\documentclass[10pt,twocolumn,letterpaper]{article}

\usepackage{cvpr}              

\usepackage{graphicx}
\usepackage{amsmath}
\usepackage{amssymb}
\usepackage{booktabs}
\usepackage{times}
\usepackage{fontawesome5}
\usepackage{multirow}
\usepackage{bbding}

\usepackage{algorithm}
\usepackage{algorithmic}

\usepackage{newfloat}
\usepackage{listings}
\DeclareCaptionStyle{ruled}{labelfont=normalfont,labelsep=colon,strut=off}
\floatstyle{ruled}
\newfloat{listing}{tb}{lst}{}
\floatname{listing}{Listing}

\newcommand{\keypoint}[1]{\vspace{0.01cm}\noindent\textbf{#1}}
\newcommand{\modelname}{4-Doodle}

\definecolor{cvprblue}{rgb}{0.21,0.49,0.74}
\usepackage[pagebackref,breaklinks,colorlinks,allcolors=cvprblue]{hyperref}

\def\paperID{*****} 
\def\confName{CVPR}
\def\confYear{2026}

\title{\modelname: Text to 3D Sketches that Move!}

\author{
Hao Chen\textsuperscript{1}\quad
Jiaqi Wang\textsuperscript{1}\quad 
Yonggang Qi\textsuperscript{1}\textsuperscript{\faEnvelope}\quad
Ke Li\textsuperscript{1}\quad
Kaiyue Pang\textsuperscript{2}\quad
Yi-Zhe Song\textsuperscript{2}\\
\textsuperscript{1}School of Artificial Intelligence, Beijing University of Posts and Telecommunications\\
\textsuperscript{2}SketchX, CVSSP, University of Surrey\\
{\tt\small \textsuperscript{\faEnvelope} Corresponding author}
}

\begin{document}
\twocolumn[{%
\renewcommand\twocolumn[1][]{#1}%
\maketitle
    

\vspace{-1cm}
\noindent
\hspace{-1.1cm}
\includegraphics[width=1.1\textwidth]{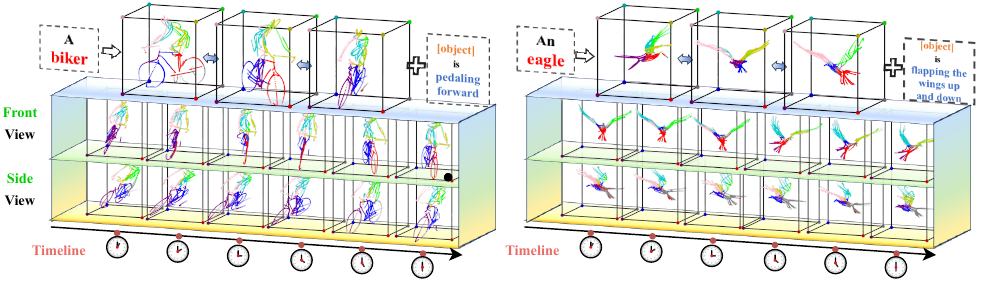}
\captionof{figure}{Given a text description, a dynamic 3D vector sketch can be generated by our model. Depth cues are colour-coded.}
\label{fig:open}
\vspace{0.5cm}
}]

\begin{abstract}
    We present a novel task: text-to-3D sketch animation, which aims to bring freeform sketches to life in dynamic 3D space. Unlike prior works focused on photorealistic content generation, we target sparse, stylized, and view-consistent 3D vector sketches, a lightweight and interpretable medium well-suited for visual communication and prototyping. However, this task is very challenging: (i) no paired dataset exists for text and 3D (or 4D) sketches; (ii) sketches require structural abstraction that is difficult to model with conventional 3D representations like NeRFs or point clouds; and (iii) animating such sketches demands temporal coherence and multi-view consistency, which current pipelines do not address. Therefore, we propose \modelname, the first training-free framework for generating dynamic 3D sketches from text. It leverages pretrained image and video diffusion models through a dual-space distillation scheme: one space captures multi-view-consistent geometry using differentiable B\'ezier curves, while the other encodes motion dynamics via temporally-aware priors. Unlike prior work (e.g., DreamFusion), which optimizes from a single view per step, our multi-view optimization ensures structural alignment and avoids view ambiguity, critical for sparse sketches. Furthermore, we introduce a structure-aware motion module that separates shape-preserving trajectories from deformation-aware changes, enabling expressive motion such as flipping, rotation, and articulated movement. Extensive experiments show that our method produces temporally realistic and structurally stable 3D sketch animations, outperforming existing baselines in both fidelity and controllability. We hope this work serves as a step toward more intuitive and accessible 4D content creation.
\end{abstract}
\section{Introduction}

Sketching has long been a universal and intuitive medium for creative expression, from prehistoric cave art to digital tablets. Now, with the rise of spatial computing platforms like Apple Vision Pro and Meta Quest, the ability to create and animate 3D sketches becomes not only desirable but foundational to immersive content creation. \emph{Can we animate what we imagine, directly from text into expressive 3D sketches that move, twist, and come alive?}

While recent breakthroughs in generative AI have enabled impressive content synthesis from natural language, including images (e.g., DALL·E 2~\cite{DALLE2}), videos (e.g., SVD~\cite{blattmann2023stable}), and even static 3D assets (e.g., DreamFusion~\cite{poole2022dreamfusion}), they remain largely grounded in photorealistic, static, or viewpoint-fixed representations. Crucially, text-driven generation of dynamic 3D sketches remains unexplored, despite its unique value in design prototyping, visual storytelling, and spatial user interfaces. 

Several recent works have begun exploring sketch-to-3D or sketch animation generation, but all come with key limitations. SketchDream~\cite{sketchdream} and Sketch2NeRF~\cite{sketch2nerf} extend 2D sketches to 3D representations using NeRFs~\cite{mildenhall2020nerfrepresentingscenesneural} and ControlNets~\cite{controlnet}, but focus on static modeling rather than animation. Sketch2Anim~\cite{zhong2025sketch2animtransferringsketchstoryboards} focuses on storyboard-to-motion transfer via key pose recovery, but assumes manually drawn motion trajectories and lacks generative motion diversity. Diff3DS~\cite{Diff3DS} pioneers differentiable B\'ezier-based 3D sketch rendering, enabling view-consistent geometry, yet remains limited to static outputs. More broadly, methods like Animate3D \cite{animate3d} and 3DTopia \cite{3dtopia} target animating photorealistic models or accelerating text-to-3D generation, but lack structural abstraction or sketch-awareness. Even multi-modal generative foundation models like CLAY \cite{zhang2024claycontrollablelargescalegenerative} offer no support for dynamic sketch abstraction or temporal consistency across views. Despite the mentioned progress, the fundamental problem (i.e., generating dynamic, spatially consistent 3D vector sketches directly from text) remains unsolved.

Tackling this task is fundamentally challenging. First, there exists no large-scale paired dataset of text descriptions and 3D (let alone 4D) vector sketches, making fully supervised training infeasible. Second, 3D sketches are inherently structural and view-dependent, requiring representations that are more abstract and lightweight than voxels, meshes, or point clouds. Third, animating sketches requires temporal coherence and spatial consistency under arbitrary viewpoint transformations, far beyond what traditional sketch animation or video synthesis can handle.

To address these challenges, we introduce the first framework for text-to-3D sketch animation, dubbed \modelname, which generates spatially coherent and dynamically animated 3D vector sketches directly from natural language. Our key insight is to decompose the problem into two interacting knowledge spaces: (i) a structure space, responsible for capturing multi-view-consistent 3D geometry using image diffusion priors, and (ii) a motion space, which encodes temporal dynamics using video diffusion priors. This dual-space formulation enables us to leverage powerful pre-trained models while bypassing the need for paired text–4D sketch data.

Unlike prior work such as DreamFusion \cite{poole2022dreamfusion}, which optimizes a single random camera view per iteration, our method aggregates gradients from multiple canonical viewpoints. This multi-view consistent optimization reduces ambiguities such as duplicate heads and enforces structural alignment across views, a crucial requirement for sketches with sparse visual cues and high semantic abstraction.

At the core of our representation is a differentiable Bézier curve-based neural sketch model, which encodes 3D curves in a resolution-independent manner. A single Bézier curve, defined by a few control points, can represent complex shapes across views (e.g., the front part depicting a horse’s head, while the back bends into a tail). Compared to NeRFs' dense sampling and heavy inference, Bézier curves are lightweight and interpretable. Point clouds, though detailed, require thousands of points and often lack the continuity and semantic clarity inherent to curves.

Based on this representation, we introduce a structure-aware motion generation module that decomposes dynamics into shape-preserving and deformation components, and a dual-space knowledge distillation scheme to transfer priors across structure and motion. This enables our model to animate sketches with fine-grained, coherent motion such as flipping, rotation, or articulated deformation.


In summary, our main contributions are as follows: (i)~We propose the first framework for generating dynamic 3D sketches from text descriptions via knowledge distillation, without requiring paired text–sketch supervision; (ii)~We design a dual-space architecture based on differentiable Bézier curve that separately models structural geometry and temporal motion while enabling cooperative reasoning between them; (iii)~We achieve state-of-the-art performance on complex spatial animations, particularly for challenging motions like rotation and flipping. Extensive experiments demonstrate that our method significantly surpasses existing approaches in terms of structural stability and motion realism.

\section{Related Work}
\begin{figure*}
    \centering
    \vspace{-0.42cm}
    \includegraphics[width=\linewidth]{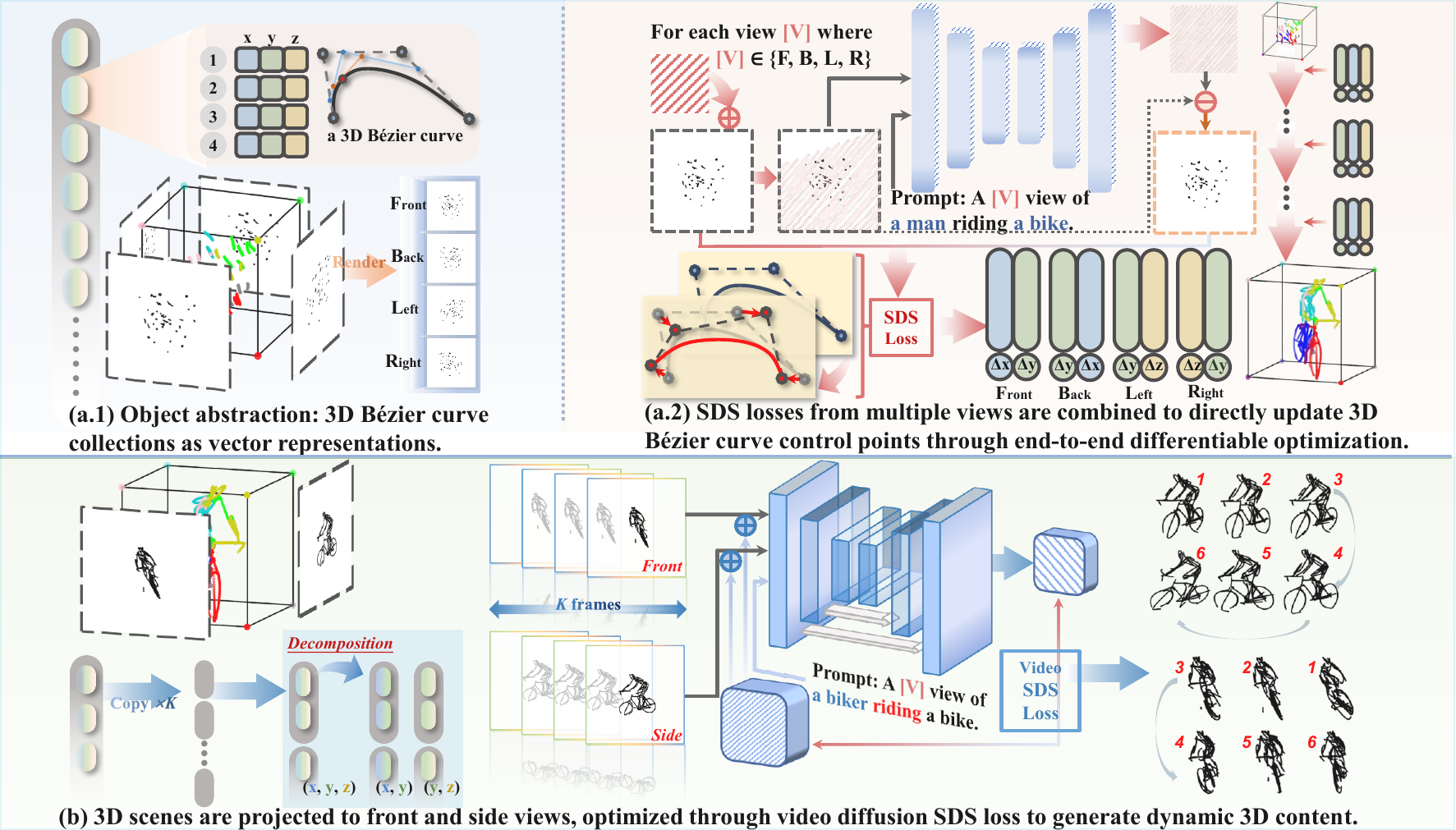}
    \caption{Overview of \modelname: text-driven dynamic 3D sketch generation. Stage I: Multi-view 3D sketch optimization using SDS loss from randomly initialized B\'ezier curves. Stage II: Motion field learning through projection-reconstruction strategy that decomposes 3D scenes into front/side views for video SDS optimization.
    }
    \vspace{-0.6cm}
    \label{fig:overview}
\end{figure*}
\keypoint{Text-to-4D Generation.}
Deep generative models have transformed content creation, evolving from 2D to 4D generation. Early breakthroughs in text-to-image synthesis (e.g., Imagen~\cite{saharia2022photorealistictexttoimagediffusionmodels}, DALL·E 2~\cite{DALLE2}) were followed by text-to-video models such as Stable Video Diffusion~\cite{blattmann2023stable}, ModelScope~\cite{modelscope}, and VideoCrafter~\cite{chen2023videocrafter1, chen2024videocrafter2}, which introduced temporal coherence. More recently, the field has advanced toward 4D generation—modeling dynamic 3D content over time. Leveraging representations like Neural Radiance Fields (NeRF)~\cite{mildenhall2020nerfrepresentingscenesneural} and 3D Gaussians (3DGS)~\cite{kerbl3Dgaussians}, models such as DreamFusion~\cite{poole2022dreamfusion} and DreamGaussian~\cite{tang2023dreamgaussian} achieve text-driven static 3D asset generation without 3D supervision via Score Distillation Sampling (SDS)~\cite{poole2022dreamfusion}. Building on this, recent works like MAV3D~\cite{singer2023text}, AYG~\cite{ling2024align}, and 4D-fy~\cite{bahmani20244d} explore dynamic 3D scene generation. However, text-to-4D vector sketch generation remains untouched—our work is the first to tackle this important yet overlooked frontier.


\keypoint{3D Sketch Generation.}
With the rise of spatial computing devices like Vision Pro and Quest, 3D sketch generation has attracted growing interest for its applications in immersive design. Early methods relied on GANs to synthesize 3D sketches~\cite{gryaditskaya2020lifting,li2023synthesizing}. More recent work, such as 3Doodle~\cite{choi20243doodle}, represents sketches as learnable parametric 3D Bézier curves, optimized with LPIPS~\cite{zhang2018unreasonableeffectivenessdeepfeatures} and CLIP~\cite{radford2021learningtransferablevisualmodels} losses from multi-view images. In contrast, our method enables text-driven 3D sketch generation without requiring explicit 3D supervision.

\keypoint{Sketch Animation.}
Sketch animation has progressed from traditional keyframe-based methods with predefined poses~\cite{bertasius2019learning}, to learning-based approaches that capture motion patterns from reference videos~\cite{bertasius2019learning} or apply dynamic deformations~\cite{jiang2012trajectory}. Other works leverage physical simulation~\cite{rohrlich1990computer} or predefined motion libraries~\cite{rusert2011selecting}. More recently, generative models—especially text-to-video diffusion—have enabled flexible, prompt-driven animation, as seen in LiveSketch~\cite{gal2024breathing}, which animates static sketches based on textual input. Unlike these methods that operate in 2D or require input sketches, our approach uniquely generates animated 3D sketches directly from text.
\section{Method}

\modelname \ introduces a novel two-stage framework for text-driven 3D dynamic sketch synthesis, as illustrated in Figure~\ref{fig:overview}. The pipeline decomposes this challenging task into two complementary stages: first, constructing coherent 3D sketch structures via multi-view consistency; then, animating these static representations by learning motion fields. Specifically, the process begins with randomly initialized 3D Bézier curves, which are progressively refined into semantically meaningful dynamic sketches. In the first stage, spatial coherence is established by optimizing curve parameters under the guidance of multi-view Score Distillation Sampling (SDS). The second stage introduces temporal dynamics through a tailored projection–reconstruction strategy that effectively incorporates video generation priors into 3D space. We describe each key module below.



\subsection{Stage I: Multi-view 3D Sketch Generation}

The 3D sketch representation is built upon parametric B\'ezier curves, where each stroke $\mathcal{S}_i$ is defined by four 3D control points $\{\mathbf{p}_{i,j} \in \mathbb{R}^3\}_{j=0}^{3}$. 
This compact parameterization captures essential geometric characteristics while maintaining differentiability for gradient-based optimization. The full 3D sketch $\mathcal{S} = \{\mathcal{S}_i\}_{i=1}^{N}$ consists of $N$ independent curves, each contributing to the overall structure.

For any viewpoint $\mathbf{v}$, the 3D sketch is rendered into a 2D image $I_{\mathbf{v}}$ via perspective projection and differentiable rasterization. Notably, 3Doodle~\cite{choi20243doodle} showed that 3D B\'ezier curves form rational B\'ezier curves after perspective projection. When the control point weights are approximately equal (i.e., when the camera is sufficiently distant from the object), this projection can be effectively approximated as standard 2D Bézier curves, thus facilitating differentiable rendering.


Specifically, our approach employs a distance field-based rasterization strategy, where the intensity at pixel $(x,y)$ is determined by contributions from all projected curves:

\begin{equation}
I_{\mathbf{v}}(x,y) = \sum_i \int w(d((x,y), \tilde{c}_{i,\mathbf{v}}(t)))dt
\end{equation}
where $d((x,y), \tilde{c}_{i,\mathbf{v}}(t))$ computes the shortest distance from pixel to curve, and $w(d) = \max(0, 1-d^2/\sigma^2)^2$ provides smooth decay. Crucially, gradients can propagate directly through the rendering process to the 3D control points:
\begin{equation}
\frac{\partial \mathcal{L}}{\partial \mathbf{p}_{i,j}} = \sum_{\mathbf{v}} \sum_{x,y} \frac{\partial \mathcal{L}}{\partial I_{\mathbf{v}}(x,y)} \cdot \frac{\partial I_{\mathbf{v}}(x,y)}{\partial \tilde{c}_{i,\mathbf{v}}} \cdot \frac{\partial \tilde{c}_{i,\mathbf{v}}}{\partial \mathbf{p}_{i,j}}
\end{equation}
This end-to-end differentiability enables direct optimization of control point parameters in 3D space, enabling effective 3D structure learning through gradient accumulation across multiple orthogonal viewpoints. In particular, our method focuses on four cardinal viewpoints $\mathbf{v} \in \{\text{front, back, left, right}\}$ to ensure comprehensive spatial coverage while maintaining computational efficiency. Formally, the optimization objective utilizes the Score Distillation Sampling (SDS), adapted for multi-view sketch generation:
\begin{equation}
\mathcal{L}_{\text{3D}} = \sum_{\mathbf{v}} \mathbb{E}_{t,\epsilon} \left[ w(t) \left\| \epsilon_\phi(z_t^{\mathbf{v}}; y_{\mathbf{v}}, t) - \epsilon \right\|_2^2 \right]
\end{equation}
where $z_t^{\mathbf{v}}$ represents the noisy latent encoding of the rendered view, $y_{\mathbf{v}}$ denotes the view-dependent text prompt ``A $\mathbf{v}$ view of [object description]'', and $w(t)$ provides time-dependent weighting. This formulation naturally encourages multi-view consistency while leveraging the rich semantic understanding of pre-trained diffusion models.

The view-dependent prompting strategy is crucial for generating geometrically coherent 3D structures. By explicitly conditioning each viewpoint on its spatial context, this strategy guides optimization toward solutions that maintain proper depth relationships and avoid the multi-face artifacts common in naive 3D synthesis approaches.

\subsection{Stage II: Motion Field Learning for 3D Animation}

The second stage transforms the static 3D sketch into dynamic sequences through learned displacement fields. Essentially, motion is parameterized as additive offsets to original control points: $\mathbf{p}_{i,j}^{(k)} = \mathbf{p}_{i,j}^{(0)} + \Delta\mathbf{p}_{i,j}^{(k)}$, where superscript $(k)$ denotes frame index and $\Delta\mathbf{p}_{i,j}^{(k)}$ represents displacement vectors. The key insight for achieving 3D motion synthesis lies in the projection-reconstruction strategy. It projects the 3D scene onto two orthogonal planes: a frontal plane capturing $(x,y)$ coordinates and a sagittal plane capturing $(y,z)$ coordinates. Each projected view generates a $K$-frame sequence, flattened into vector representations:

\begin{align}
\mathbf{v}_{\text{front}}^{(k)} &= \text{flatten}(\{(x_{i,j}^{(k)}, y_{i,j}^{(k)})\}_{i,j}) \\
\mathbf{v}_{\text{side}}^{(k)} &= \text{flatten}(\{(y_{i,j}^{(k)}, z_{i,j}^{(k)})\}_{i,j})
\end{align}
The flattened representations serve as compact encodings of temporal evolution for each view. Video generation models process these sequences to predict displacement patterns consistent with natural motion dynamics described in text prompts. By leveraging video SDS loss, the model predicts corresponding displacement sequences $\Delta\mathbf{v}_{\text{front}}^{(k)}$ and $\Delta\mathbf{v}_{\text{side}}^{(k)}$ for each view.

The key innovation lies in how two 2D displacement sequences are recombined into complete 3D displacement vectors. From frontal view displacement $\Delta\mathbf{v}_{\text{front}}^{(k)}$, displacement components in $x$ and $y$ directions can be directly obtained. From lateral view displacement $\Delta\mathbf{v}_{\text{side}}^{(k)}$, displacement components in $y$ and $z$ directions can be obtained. For tackling $y$-coordinate discrepancies, an averaging strategy is adopted:

\begin{align}
\Delta x_{i,j}^{(k)} &= \Delta\mathbf{v}_{\text{front}}^{(k)}[2(i \cdot 4 + j)] \\
\Delta y_{i,j}^{(k)} &= \frac{\Delta\mathbf{v}_{\text{front}}^{(k)}[2(i \cdot 4 + j) + 1] + \Delta\mathbf{v}_{\text{side}}^{(k)}[2(i \cdot 4 + j)]}{2} \\
\Delta z_{i,j}^{(k)} &= \Delta\mathbf{v}_{\text{side}}^{(k)}[2(i \cdot 4 + j) + 1]
\end{align}
To this end, the full 3D displacement vector is obtained:
\begin{equation}
    \Delta\mathbf{p}_{i,j}^{(k)} = [\Delta x_{i,j}^{(k)}, \Delta y_{i,j}^{(k)}, \Delta z_{i,j}^{(k)}]^T
\end{equation}
Building on recent advances in sketch animation~\cite{gal2024breathing}, our method adapts motion prior distillation to 3D settings. The key distinction is its ability to synthesize motion in 3D space, rather than being limited to 2D planar deformations. This extension greatly enhances the expressiveness of the generated animations while preserving the semantic richness of video generation models.

\subsection{Loss Function and Optimization}

The training objective employs a staged optimization strategy, achieving multi-view spatial consistency and temporal motion coherence through carefully designed loss functions at different stages. Given the fundamentally different optimization goals and constraints of the two stages, sequential rather than simultaneous optimization ensures each stage focuses on its core objectives.

\textbf{First Stage Optimization} focuses on establishing robust 3D structures, with loss functions combining semantic alignment and geometric constraints:

\begin{equation}
\mathcal{L}_{\text{stage1}} = \mathcal{L}_{\text{3D}} + \lambda_g \mathcal{L}_{\text{geometric}}
\end{equation}
where the multi-view SDS loss is as previously described. To maintain structural integrity, a geometric consistency loss constrains relative configurations between control points:

\begin{equation}
\mathcal{L}_{\text{geometric}} = \frac{1}{N} \sum_{i=1}^{N} \sum_{j=0}^{2} \left\|\frac{\mathbf{p}_{i,j+1} - \mathbf{p}_{i,j}}{\|\mathbf{p}_{i,j+1} - \mathbf{p}_{i,j}\|} - \frac{\mathbf{p}_{i,j} - \mathbf{p}_{i,j-1}}{\|\mathbf{p}_{i,j} - \mathbf{p}_{i,j-1}\|}\right\|_2^2
\end{equation}
This loss function preserves topological structure by constraining the normalized direction vectors between adjacent control points, preventing unrealistic geometric deformations during optimization. The first stage of optimization continues until the 3D sketch structure converges, ensuring that the generated static sketches exhibit strong multi-view consistency and geometric coherence.

\begin{figure*}
    \centering
    \includegraphics[width=\linewidth]{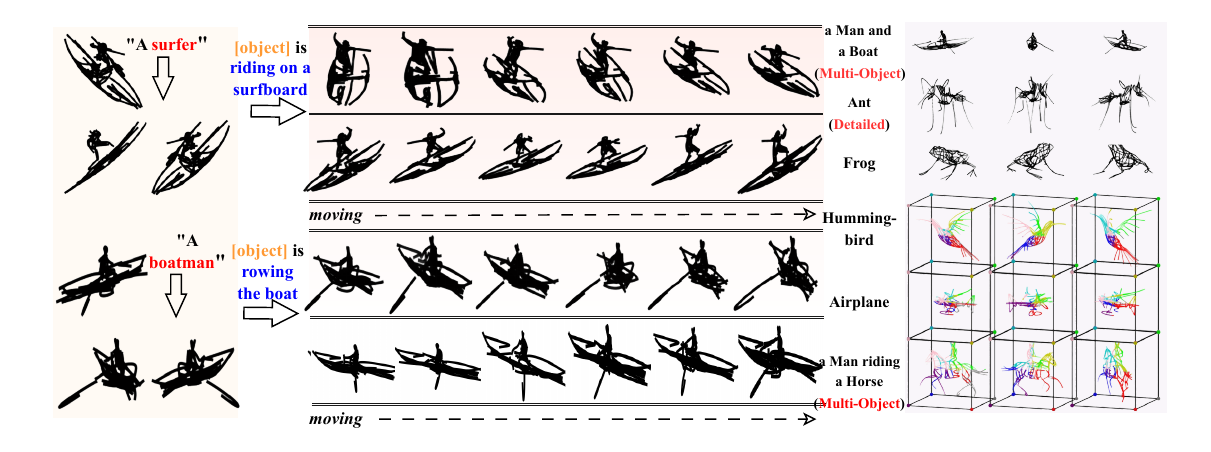}
    \vspace{-0.5cm}
    \caption{Text-to-4D Sketch Generation. Generated 4D sketches from text prompts showing motion sequences and multi-object scenes with temporal dynamics and sketch-style representation.
    }
    \vspace{-0.6cm}
    \label{fig:expshow}
\end{figure*}

\textbf{Second Stage Optimization} introduces motion dynamics based on fixed 3D structures, with loss functions leveraging video diffusion priors:

\begin{equation}
\mathcal{L}_{\text{motion}} = \mathcal{L}_{\text{video}}^{\text{front}} + \mathcal{L}_{\text{video}}^{\text{side}}
\end{equation}
where each video loss term follows standard SDS formulation applied to temporal sequences:

\begin{equation}
\mathcal{L}_{\text{video}}^{\text{view}} = \mathbb{E}_{t,\epsilon} \left[ w(t) \left\| \epsilon_\psi(\mathbf{z}_t^{\text{view}}; y_{\text{motion}}, t) - \epsilon \right\|_2^2 \right]
\end{equation}
where $\mathbf{z}_t^{\text{view}}$ represents noisy encoding of flattened motion sequences, $y_{\text{motion}}$ denotes motion description prompts, and $\epsilon_\psi$ is the noise prediction network of the video diffusion model.

Through this staged optimization strategy, interference between structure formation and motion learning is avoided. The solid 3D structure established in the first stage provides a reliable foundation for motion learning in the second stage, while the second stage introduces rich temporal dynamics while maintaining structural integrity.

To ensure smooth temporal transitions, additional regularization terms are incorporated in the second stage to penalize sudden changes in displacement magnitude:

\begin{equation}
\mathcal{L}_{\text{smooth}} = \sum_{i,j,k} \left\| \Delta\mathbf{p}_{i,j}^{(k+1)} - \Delta\mathbf{p}_{i,j}^{(k)} \right\|_2^2
\end{equation}
which is important for generating visually pleasing animations, avoiding jittery or unrealistic motion patterns. The complete framework achieves organic unity of semantic alignment, geometric consistency, and temporal smoothness through staged optimization, producing high-quality 3D dynamic sketches that faithfully reflect input text descriptions.

Based on curriculum learning principles, the staged optimization framework achieves coarse-to-fine feature learning through progressive knowledge distillation. Let the complete parameter space of the 3D sketch be $\Theta = \{\mathbf{p}_{i,j}\}_{i=1,j=0}^{N,3}$, where $N$ is the number of curves. In the first stage, an adaptive timestep scheduling strategy $t \sim \mathcal{U}[t_{\max}, t_{\min}]$ is employed, where $t_{\max}$ linearly decays from 0.8 to 0.6, while $t_{\min}$ remains at 0.02. High noise timesteps ($t > 0.5$) primarily handle global semantic alignment, guiding structure formation; low noise timesteps ($t < 0.3$) focus on detail refinement, improving visual quality. Second stage displacement field learning follows similar progressive strategies but with special design for temporal consistency. Displacement magnitude is modulated through dynamic weights $\alpha(t) = 1 - \exp(-\beta t)$, ensuring moderate motion amplitude initially with gradually enhanced expressiveness later.

\section{Experiments}
Since there are no existing prior works on text-to-4D sketch generation, we alternatively evaluate our model from two perspectives, i.e., the results of text-to-3D sketch generation and motion generation, respectively.
\label{sec:4}

\subsection{Experimental Settings}
\label{sec:4.1}

\subsubsection{Implementation Details}
\keypoint{Text-to-3D Sketch Generation.} 
By default, our model employs 16 cubic Bézier curves (4 control points) to construct a 3D sketch.
To ensure continuity and diversity in 3D space, we devise a hierarchical initialization scheme, i.e., we first randomly sample initial control points within a spherical region of radius 0.2, followed by gradually constructing subsequent control points by increasing minimum distance constraints (0.001) and random displacement (maximum 0.01) based on the previous control point. During rasterization, we render strokes in black with a fixed width. Adam optimizer is adopted with a learning rate set to 1.5e-3 throughout 4000 training iterations. Default CFG weight $w_s$ is set to 7.5. Text-to-image model, Stable Diffusion 2.1, is used for static 3D structure generation. We consistently use four basic viewpoints: front, left, right, and back views. Additionally, top-view supervision is used during training with a probability of $10\%$, providing extra benefits for the overall 3D sketch generation.


\keypoint{3D Motion Generation.} We employ a lightweight {MLP to parameterize the transformation functions.} We find out that setting CFG weight $w_m$ larger, i.e., 30, achieves better results. The same Adam optimizer is utilized with a fixed learning rate of 1e-4 during training. Notably, for efficiency, a shared MLP network is employed to simultaneously process 3D sketch projections from both front and side views. ModelScope \cite{modelscope} is employed as the motion prior. The optimization iteration steps are set to 1K.


\keypoint{Conditional Texts} {We adopt the same text descriptions} from LiveSketch \cite{gal2024breathing} as text inputs, which cover diverse activities of humans, animals, and common objects (details in the Supplemental). 

\begin{table}[t]
    \centering
    \begin{tabular}{l|ccc}
        \toprule
       CLIP (T2I)  & MVDream & DiffSketcher & \textbf{Ours} \\
       \hline
       Front view &0.312&0.257&\textbf{0.316}\\
       Left view &\textbf{0.308}&0.277&0.307\\
       Back view &0.309&0.253&\textbf{0.312}\\
       Right view &0.303&0.248&\textbf{0.322}\\
       Average &0.308&0.260&\textbf{0.314}\\
       \bottomrule
    \end{tabular}
    \caption{Text-to-3D generation results of CLIP score between input text and 2D sketches rendered from our and baseline methods.}
    \label{tab:t2-3d}
\end{table}

\begin{figure*}[t]
    \centering
    \includegraphics[width=0.98\linewidth]{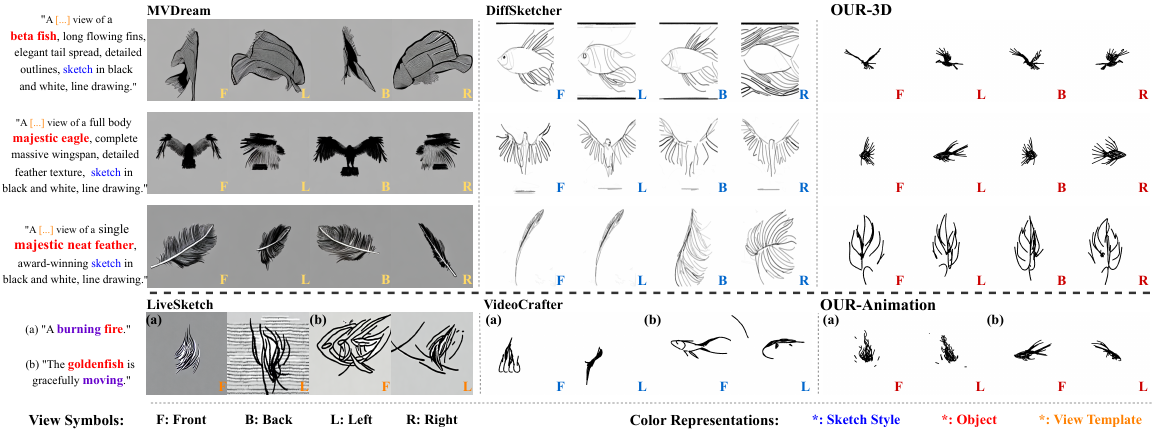}
    \vspace{-0.2cm}
    \caption{Comparison results of text-to-3D (in different viewpoints) and sketch animation.}
    \vspace{-0.4cm}
    \label{fig:text-to-3d-comp}
\end{figure*}

\subsubsection{Baseline Methods} 
\keypoint{Text-to-3D Sketch Generation.} Due to the absence of available text-to-3D generation baseline methods, we compare with state-of-the-art text-to-2D sketch generator DiffSketcher \cite{xing2023diffsketcher} {using our projected 2D sketches from different views.} Notably, to facilitate comparison in a specific viewpoint, we append a textual description of the viewpoint, e.g., ``front view," to the input text prompt to DiffSketcher. Moreover, 
a strong novel view synthesis method MVDream \cite{shi2023mvdream} is also compared. Similarly, to enforce the model to generate sketches in a specific viewpoint, additional textual descriptions, e.g., ``black and white line drawing, front view", are also provided to indicate the desired views and image style. 



\keypoint{3D Motion Generation.} To gauge the motion quality of the generated 3D sketch, we compare with LiveSketch \cite{gal2024breathing} which is state-of-the-art text-to-2D sketch animation approach. Note that unlike our model which only needs a text description as input, LiveSketch requires an additional input of vector sketch. Therefore, we feed LiveSketch with the 2D sketch projected from the 3D sketch generated by our model for a fair comparison. {In addition, we compare with a powerful text-to-video generation model, i.e., VideoCrafter \cite{chen2024videocrafter2,chen2023videocrafter1}.} The same textual prompts are used for all models.


\begin{table}[t]
    \centering
    \vspace{-0.2cm}
    \begin{tabular}{l|cc|cc}
       \toprule
       \multirow{2}{*}{Method}  & \multicolumn{2}{c}{CLIP (I2I) score}  & \multicolumn{2}{c}{X-CLIP score} \\
                                & front & side  & front & side \\
       \hline
       Ours & \underline{0.896} & \underline{0.897}   & 0.146  & 0.156 \\
       LiveSketch & \textbf{0.918} & \textbf{0.930}       & \underline{0.177} &   \textbf{0.163}\\
       VideoCrafter & 0.846 & 0.870     & \textbf{0.196}   &  \underline{0.160}\\
       \bottomrule
    \end{tabular}
    \caption{Motion quality comparison results.}
    \vspace{-0.8cm}
    \label{tab:t2-vedio}
    
\end{table}

\subsection{Results}
\subsubsection{Text-to-3D Sketch Generation}
\label{sec:4.2}
\keypoint{Quantitative Results.} Following LiveSketch \cite{gal2024breathing}, CLIP (T2I) score is adopted to measure the similarity between the input text and the 2D sketch images. The sketch images are either from the projections of the generated 3D sketch by our model or 2D sketches directly generated by MVDream and DiffSketcher. As shown in Table~\ref{tab:t2-3d}, our model can generate sketches that can better match the input texts at different viewpoints.


\keypoint{Qualitative Analysis.} Some exemplar results from different viewpoints {of 3D sketches are shown in Figure~\ref{fig:expshow}. We can see that the 2D projections from our generated 3D sketches exhibit clear geometry details in different viewpoints. Comparison results are shown in Figure~\ref{fig:text-to-3d-comp}, we observe significant limitations in baseline methods regarding sketch-style content generation. Even with explicit sketch style specifications in the prompts, these methods struggle to produce satisfactory line-based expressions. In contrast, our method not only accurately captures the artistic characteristics of sketches but also achieves style uniformity while maintaining a reasonable 3D structure.



\subsubsection{3D Motion Generation}
\label{sec:4.3}

\begin{figure*}
    \centering
    \includegraphics[width=\linewidth]{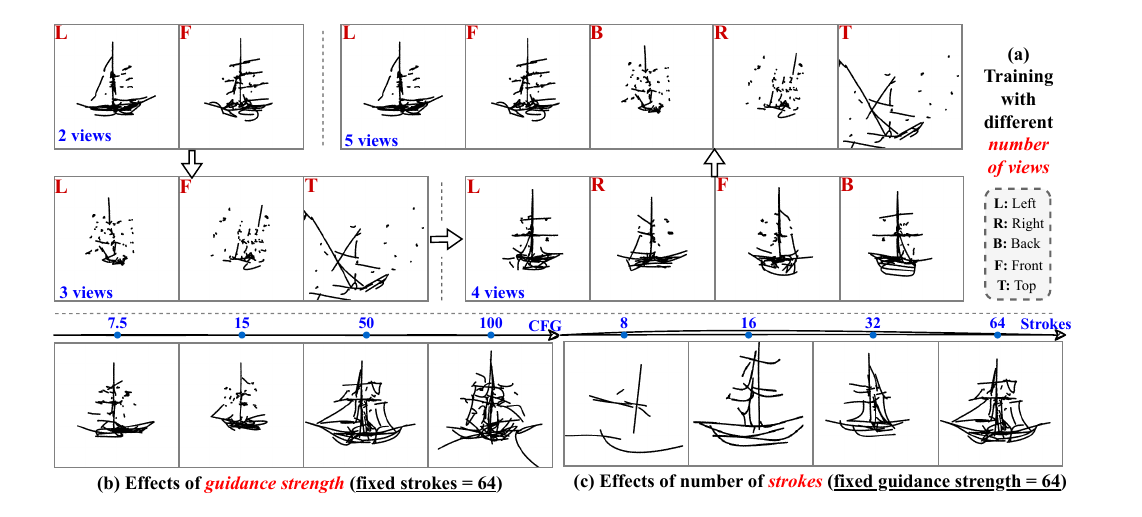}
    \vspace{-1cm}
    \caption{Training with different number of views and effects of guidance strength and number of strokes.}
    \label{fig:view-ablation}
    \vspace{-0.6cm}
\end{figure*}

\begin{figure}[htbp]
    \centering
    \includegraphics[width=\linewidth]{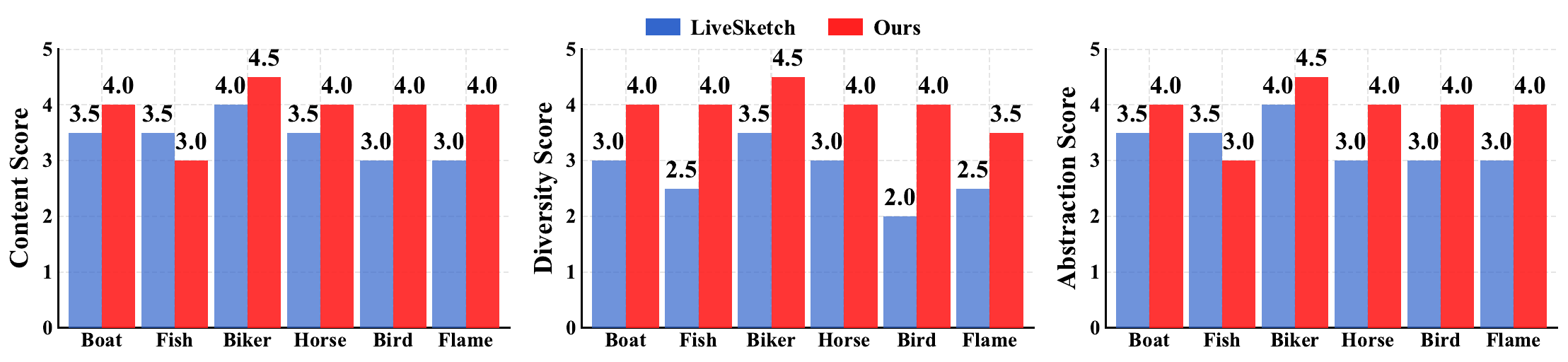}
    \vspace{-0.6cm}
    \caption{Quality evaluation scores from Qwen-VLM comparing LiveSketch and our approach across different objects.}
    \vspace{-0.6cm}
    \label{fig:comp_bar}
\end{figure}

\keypoint{Quantitative Results.} Similarly, as there is no existing measurement for evaluating dynamic 3D sketches, we opt for measuring the dynamic 3D sketch using its 2D motion projections from different viewpoints. Following LiveSketch \cite{gal2024breathing}, there are two metrics employed to evaluate the motion quality of the projected 3D sketches by our model.
First, {given a specific view angle}, we measure the ``sketch-to-video" consistency, i.e., CLIP (I2I), by computing the average cosine similarity between video frames and input sketches using their CLIP features. Second, we introduce the X-CLIP evaluation metric (a model extending CLIP to video recognition) to assess the semantic alignment between generated videos and text prompts, i.e., ``text-to-video" consistency. Results in Table~\ref{tab:t2-vedio} show that ours can achieve comparative results over baseline methods. However, we found that the CLIP and X-CLIP scores are not reliable measurements for 3D sketches due to the limitations on reflecting structural expressiveness, appearance refinement, and view consistency (please refer to the supplementary material for more details). Therefore, we employ an open-source vision-language model (VLM), i.e., Qwen2-VL-7B, for further evaluation.

\noindent \emph{Quality assessment using VLM.} Utilizing Qwen2-VL-7B, we measure the appearance quality of the 2D motion projections from the dynamic 3D sketch. Specifically, a text prompt is devised to require Qwen2-VL-7B to rate (scores 1 to 5) these 2D projections in terms of content completeness, diversity, and abstraction level (more details in the supplementary material). As shown in Figure~\ref{fig:comp_bar}, ours can clearly outperform competitors.
\vspace{-0.2cm}


\paragraph{Qualitative Analysis.} 
The lower panel of Figure~\ref{fig:text-to-3d-comp} demonstrates representative frames from output videos in the front and side views given the same conditional text prompts. VideoCrafter is clearly suffering from noticeable visual artifacts and struggle to maintain consistent visual style. As the pioneering work in 2D dynamic sketch generation, LiveSketch demonstrates convincing performances in dynamic modelling, yet limited in handling different viewpoints, e.g., side views in this case. In comparison, our proposed method achieves significant improvements in both generation stability and quality. More video demos are provided in the supplementary materials. 

\subsection{Ablation Study}
\label{sec:4.4}

\keypoint{Supervision views for text-to-3D generation.} Our model is trained using text-to-image diffusion models with a set of discrete viewpoints for 3D construction. We therefore conduct experiments by using different view settings with other hyperparameters fixed.
Moreover, as shown in Figure~\ref{fig:view-ablation} (a), the qualitative results reveal that (i) using more views is better, and (ii) the top view could taint the overall performance; we attribute this issue to the unreliable top-view prior knowledge from the image diffusion model probably due to the lack of training data.

\keypoint{Guidance strength of conditional text.} To further verify the impact of guidance strength during 3D construction, we gradually increase its value from 7.5 to measure the performance. As shown in Figure~\ref{fig:view-ablation} (b), the fidelity of the generation is improved using a larger guidance strength, while too large a value will result in a noisy visual appearance. 

\keypoint{Number of strokes.} To inspect the influence of stroke numbers, we adopt different choices. As shown in Figure~\ref{fig:view-ablation} (c), setting a proper number of strokes is crucial to achieve visually appealing results without losing important details or being too complex.
\vspace{-0.2cm}



\section{Conclusion}

In this work, we present a novel approach for text-driven dynamic 3D vector sketch generation, extending traditional sketch generation into the domain of animated 3D content. Central to our method is a dual-space knowledge distillation framework, which leverages pre-trained image and video diffusion models to transfer knowledge of 3D structure and motion dynamics, eliminating the need for 3D or motion-specific training data. We demonstrate the effectiveness of this knowledge transfer between structure and motion spaces. Experimental results validate that our model can generate stable, realistic, and naturally animated 3D sketches. Furthermore, our findings highlight a gap in the current CLIP-based metrics, which may not fully capture the quality of 3D sketch generation or dynamic motion, suggesting the need for new, specialized evaluation metrics in this area. We hope this work opens new avenues for expressive and intuitive 4D content creation via text and 3D sketch.



{
    \small
    \bibliographystyle{ieeenat_fullname}
    \bibliography{main}

\begin{thebibliography}{33}
\providecommand{\natexlab}[1]{#1}
\providecommand{\url}[1]{\texttt{#1}}
\expandafter\ifx\csname urlstyle\endcsname\relax
  \providecommand{\doi}[1]{doi: #1}\else
  \providecommand{\doi}{doi: \begingroup \urlstyle{rm}\Url}\fi

\bibitem[Bahmani et~al.(2024)Bahmani, Skorokhodov, Rong, Wetzstein, Guibas, Wonka, Tulyakov, Park, Tagliasacchi, and Lindell]{bahmani20244d}
Sherwin Bahmani, Ivan Skorokhodov, Victor Rong, Gordon Wetzstein, Leonidas Guibas, Peter Wonka, Sergey Tulyakov, Jeong~Joon Park, Andrea Tagliasacchi, and David~B Lindell.
\newblock {4D-fy}: Text-to-4d generation using hybrid score distillation sampling.
\newblock In \emph{Proceedings of the IEEE/CVF Conference on Computer Vision and Pattern Recognition}, pages 7996--8006, 2024.

\bibitem[Bertasius et~al.(2019)Bertasius, Feichtenhofer, Tran, Shi, and Torresani]{bertasius2019learning}
Gedas Bertasius, Christoph Feichtenhofer, Du Tran, Jianbo Shi, and Lorenzo Torresani.
\newblock Learning temporal pose estimation from sparsely-labeled videos.
\newblock \emph{Advances in Neural Information Processing Systems}, 32, 2019.

\bibitem[Blattmann et~al.(2023)Blattmann, Dockhorn, Kulal, Mendelevitch, Kilian, Lorenz, Levi, English, Voleti, Letts, et~al.]{blattmann2023stable}
Andreas Blattmann, Tim Dockhorn, Sumith Kulal, Daniel Mendelevitch, Maciej Kilian, Dominik Lorenz, Yam Levi, Zion English, Vikram Voleti, Adam Letts, et~al.
\newblock Stable video diffusion: Scaling latent video diffusion models to large datasets.
\newblock \emph{arXiv preprint arXiv:2311.15127}, 2023.

\bibitem[Chen et~al.(2023)Chen, Xia, He, Zhang, Cun, Yang, Xing, Liu, Chen, Wang, Weng, and Shan]{chen2023videocrafter1}
Haoxin Chen, Menghan Xia, Yingqing He, Yong Zhang, Xiaodong Cun, Shaoshu Yang, Jinbo Xing, Yaofang Liu, Qifeng Chen, Xintao Wang, Chao Weng, and Ying Shan.
\newblock Videocrafter1: Open diffusion models for high-quality video generation, 2023.

\bibitem[Chen et~al.(2024{\natexlab{a}})Chen, Zhang, Cun, Xia, Wang, Weng, and Shan]{chen2024videocrafter2}
Haoxin Chen, Yong Zhang, Xiaodong Cun, Menghan Xia, Xintao Wang, Chao Weng, and Ying Shan.
\newblock Videocrafter2: Overcoming data limitations for high-quality video diffusion models, 2024{\natexlab{a}}.

\bibitem[Chen et~al.(2024{\natexlab{b}})Chen, Yuan, Wang, Sheng, He, Dong, Bo, and Guo]{sketch2nerf}
Minglin Chen, Weihao Yuan, Yukun Wang, Zhe Sheng, Yisheng He, Zilong Dong, Liefeng Bo, and Yulan Guo.
\newblock {Sketch2NeRF}: Multi-view sketch-guided text-to-{3D} generation, 2024{\natexlab{b}}.

\bibitem[Choi et~al.(2024)Choi, Lee, Park, and Kim]{choi20243doodle}
Changwoon Choi, Jaeah Lee, Jaesik Park, and Young~Min Kim.
\newblock {3Doodle}: Compact abstraction of objects with {3D} strokes.
\newblock \emph{ACM Transactions on Graphics (TOG)}, 43\penalty0 (4):\penalty0 1--13, 2024.

\bibitem[Gal et~al.(2024)Gal, Vinker, Alaluf, Bermano, Cohen-Or, Shamir, and Chechik]{gal2024breathing}
Rinon Gal, Yael Vinker, Yuval Alaluf, Amit Bermano, Daniel Cohen-Or, Ariel Shamir, and Gal Chechik.
\newblock Breathing life into sketches using text-to-video priors.
\newblock In \emph{Proceedings of the IEEE/CVF Conference on Computer Vision and Pattern Recognition}, pages 4325--4336, 2024.

\bibitem[Gryaditskaya et~al.(2020)Gryaditskaya, H{\"a}hnlein, Liu, Sheffer, and Bousseau]{gryaditskaya2020lifting}
Yulia Gryaditskaya, Felix H{\"a}hnlein, Chenxi Liu, Alla Sheffer, and Adrien Bousseau.
\newblock Lifting freehand concept sketches into {3D}.
\newblock \emph{ACM Transactions on Graphics (TOG)}, 39\penalty0 (6):\penalty0 1--16, 2020.

\bibitem[Hong et~al.(2024)Hong, Tang, Cao, Shi, Wu, Chen, Yang, Wang, Pan, Lin, and Liu]{3dtopia}
Fangzhou Hong, Jiaxiang Tang, Ziang Cao, Min Shi, Tong Wu, Zhaoxi Chen, Shuai Yang, Tengfei Wang, Liang Pan, Dahua Lin, and Ziwei Liu.
\newblock {3DTopia}: Large text-to-3d generation model with hybrid diffusion priors, 2024.

\bibitem[Jiang et~al.(2024)Jiang, Yu, Cao, Wang, Hu, and Gao]{animate3d}
Yanqin Jiang, Chaohui Yu, Chenjie Cao, Fan Wang, Weiming Hu, and Jin Gao.
\newblock {Animate3D}: Animating any {3D} model with multi-view video diffusion, 2024.

\bibitem[Jiang et~al.(2012)Jiang, Dai, Xue, Liu, and Ngo]{jiang2012trajectory}
Yu-Gang Jiang, Qi Dai, Xiangyang Xue, Wei Liu, and Chong-Wah Ngo.
\newblock Trajectory-based modeling of human actions with motion reference points.
\newblock In \emph{Computer Vision--ECCV 2012: 12th European Conference on Computer Vision, Florence, Italy, October 7-13, 2012, Proceedings, Part V 12}, pages 425--438. Springer, 2012.

\bibitem[Kerbl et~al.(2023)Kerbl, Kopanas, Leimk{\"u}hler, and Drettakis]{kerbl3Dgaussians}
Bernhard Kerbl, Georgios Kopanas, Thomas Leimk{\"u}hler, and George Drettakis.
\newblock {3D} gaussian splatting for real-time radiance field rendering.
\newblock \emph{ACM Transactions on Graphics}, 42\penalty0 (4), 2023.

\bibitem[Li(2023)]{li2023synthesizing}
Wanwan Li.
\newblock Synthesizing {3D} {VR} sketch using generative adversarial neural network.
\newblock In \emph{Proceedings of the 2023 7th International Conference on Big Data and Internet of Things}, pages 122--128, 2023.

\bibitem[Ling et~al.(2024)Ling, Kim, Torralba, Fidler, and Kreis]{ling2024align}
Huan Ling, Seung~Wook Kim, Antonio Torralba, Sanja Fidler, and Karsten Kreis.
\newblock Align your gaussians: Text-to-4d with dynamic {3D} gaussians and composed diffusion models.
\newblock In \emph{Proceedings of the IEEE/CVF Conference on Computer Vision and Pattern Recognition}, pages 8576--8588, 2024.

\bibitem[Liu et~al.(2024)Liu, Fu, Lai, and Gao]{sketchdream}
Feng-Lin Liu, Hongbo Fu, Yu-Kun Lai, and Lin Gao.
\newblock {SketchDream}: Sketch-based text-to-{3D} generation and editing, 2024.

\bibitem[Mildenhall et~al.(2020)Mildenhall, Srinivasan, Tancik, Barron, Ramamoorthi, and Ng]{mildenhall2020nerfrepresentingscenesneural}
Ben Mildenhall, Pratul~P. Srinivasan, Matthew Tancik, Jonathan~T. Barron, Ravi Ramamoorthi, and Ren Ng.
\newblock {NeRF}: Representing scenes as neural radiance fields for view synthesis, 2020.

\bibitem[Poole et~al.(2022)Poole, Jain, Barron, and Mildenhall]{poole2022dreamfusion}
Ben Poole, Ajay Jain, Jonathan~T Barron, and Ben Mildenhall.
\newblock Dreamfusion: Text-to-3d using 2d diffusion.
\newblock \emph{arXiv preprint arXiv:2209.14988}, 2022.

\bibitem[Radford et~al.(2021)Radford, Kim, Hallacy, Ramesh, Goh, Agarwal, Sastry, Askell, Mishkin, Clark, Krueger, and Sutskever]{radford2021learningtransferablevisualmodels}
Alec Radford, Jong~Wook Kim, Chris Hallacy, Aditya Ramesh, Gabriel Goh, Sandhini Agarwal, Girish Sastry, Amanda Askell, Pamela Mishkin, Jack Clark, Gretchen Krueger, and Ilya Sutskever.
\newblock Learning transferable visual models from natural language supervision, 2021.

\bibitem[Ramesh et~al.(2022)Ramesh, Dhariwal, Nichol, Chu, and Chen]{DALLE2}
Aditya Ramesh, Prafulla Dhariwal, Alex Nichol, Casey Chu, and Mark Chen.
\newblock Hierarchical text-conditional image generation with clip latents, 2022.

\bibitem[Rohrlich(1990)]{rohrlich1990computer}
Fritz Rohrlich.
\newblock Computer simulation in the physical sciences.
\newblock In \emph{PSA: Proceedings of the Biennial Meeting of the Philosophy of Science Association}, pages 507--518. Cambridge University Press, 1990.

\bibitem[Rusert et~al.(2011)Rusert, Str{\"o}m, Andersson, Wennersten, and Sj{\"o}berg]{rusert2011selecting}
Thomas Rusert, Jacob Str{\"o}m, Kenneth Andersson, Per Wennersten, and Rickard Sj{\"o}berg.
\newblock Selecting predicted motion vector candidates, 2011.
\newblock US Patent App. 13/022,170.

\bibitem[Saharia et~al.(2022)Saharia, Chan, Saxena, Li, Whang, Denton, Ghasemipour, Ayan, Mahdavi, Lopes, Salimans, Ho, Fleet, and Norouzi]{saharia2022photorealistictexttoimagediffusionmodels}
Chitwan Saharia, William Chan, Saurabh Saxena, Lala Li, Jay Whang, Emily Denton, Seyed Kamyar~Seyed Ghasemipour, Burcu~Karagol Ayan, S.~Sara Mahdavi, Rapha~Gontijo Lopes, Tim Salimans, Jonathan Ho, David~J Fleet, and Mohammad Norouzi.
\newblock Photorealistic text-to-image diffusion models with deep language understanding, 2022.

\bibitem[Shi et~al.(2023)Shi, Wang, Ye, Long, Li, and Yang]{shi2023mvdream}
Yichun Shi, Peng Wang, Jianglong Ye, Mai Long, Kejie Li, and Xiao Yang.
\newblock {MVDream}: Multi-view diffusion for 3d generation.
\newblock \emph{arXiv preprint arXiv:2308.16512}, 2023.

\bibitem[Singer et~al.(2023)Singer, Sheynin, Polyak, Ashual, Makarov, Kokkinos, Goyal, Vedaldi, Parikh, Johnson, et~al.]{singer2023text}
Uriel Singer, Shelly Sheynin, Adam Polyak, Oron Ashual, Iurii Makarov, Filippos Kokkinos, Naman Goyal, Andrea Vedaldi, Devi Parikh, Justin Johnson, et~al.
\newblock Text-to-4d dynamic scene generation.
\newblock \emph{arXiv preprint arXiv:2301.11280}, 2023.

\bibitem[Tang et~al.(2023)Tang, Ren, Zhou, Liu, and Zeng]{tang2023dreamgaussian}
Jiaxiang Tang, Jiawei Ren, Hang Zhou, Ziwei Liu, and Gang Zeng.
\newblock Dreamgaussian: Generative gaussian splatting for efficient 3d content creation.
\newblock \emph{arXiv preprint arXiv:2309.16653}, 2023.

\bibitem[Team(2023)]{modelscope}
The~ModelScope Team.
\newblock {ModelScope}: Bring the notion of model-as-a-service to life.
\newblock \url{https://github.com/modelscope/modelscope}, 2023.

\bibitem[Xing et~al.(2023)Xing, Wang, Zhou, Zhang, Yu, and Xu]{xing2023diffsketcher}
Ximing Xing, Chuang Wang, Haitao Zhou, Jing Zhang, Qian Yu, and Dong Xu.
\newblock {DiffSketcher}: Text guided vector sketch synthesis through latent diffusion models.
\newblock \emph{Advances in Neural Information Processing Systems}, 36:\penalty0 15869--15889, 2023.

\bibitem[Zhang et~al.(2023)Zhang, Rao, and Agrawala]{controlnet}
Lvmin Zhang, Anyi Rao, and Maneesh Agrawala.
\newblock Adding conditional control to text-to-image diffusion models, 2023.

\bibitem[Zhang et~al.(2024{\natexlab{a}})Zhang, Wang, Zhang, Qiu, Pang, Jiang, Yang, Xu, and Yu]{zhang2024claycontrollablelargescalegenerative}
Longwen Zhang, Ziyu Wang, Qixuan Zhang, Qiwei Qiu, Anqi Pang, Haoran Jiang, Wei Yang, Lan Xu, and Jingyi Yu.
\newblock {CLAY}: A controllable large-scale generative model for creating high-quality 3d assets, 2024{\natexlab{a}}.

\bibitem[Zhang et~al.(2018)Zhang, Isola, Efros, Shechtman, and Wang]{zhang2018unreasonableeffectivenessdeepfeatures}
Richard Zhang, Phillip Isola, Alexei~A. Efros, Eli Shechtman, and Oliver Wang.
\newblock The unreasonable effectiveness of deep features as a perceptual metric, 2018.

\bibitem[Zhang et~al.(2024{\natexlab{b}})Zhang, Wang, Zou, Wu, and Ma]{Diff3DS}
Yibo Zhang, Lihong Wang, Changqing Zou, Tieru Wu, and Rui Ma.
\newblock {Diff3DS}: Generating view-consistent {3D} sketch via differentiable curve rendering.
\newblock \emph{arXiv preprint arXiv:2405.15305}, 2024{\natexlab{b}}.

\bibitem[Zhong et~al.(2025)Zhong, Guo, Xie, Wang, and Li]{zhong2025sketch2animtransferringsketchstoryboards}
Lei Zhong, Chuan Guo, Yiming Xie, Jiawei Wang, and Changjian Li.
\newblock {Sketch2Anim}: Towards transferring sketch storyboards into {3D} animation, 2025.

\end{thebibliography}
}

\clearpage
\setcounter{page}{1}
\vspace{20pt}
\maketitlesupplementary
\vspace{20pt}

\section{Pseudocode}

Our 4-Doodle framework decomposes text-to-4D sketch generation into two sequential stages. Stage I establishes view-consistent 3D geometry through multi-view Score Distillation Sampling (SDS), while Stage II learns temporal dynamics via projection-reconstruction strategy that leverages video diffusion priors. This decomposition enables training-free generation by distilling knowledge from pretrained image and video diffusion models without requiring paired text-4D sketch datasets.

The key insight lies in representing 3D sketches as collections of parametric Bézier curves, which provide compact, interpretable, and differentiable primitives. Unlike dense representations such as NeRFs or point clouds, Bézier curves naturally capture the structural essence of sketches while maintaining computational efficiency and editability.
\subsection{Stage I: Multi-view 3D Sketch Generation}

\begin{algorithm}
\caption{Stage I: Multi-view 3D Sketch Generation}
\begin{algorithmic}[1]
\REQUIRE Text prompt $y$, number of curves $N$, iterations $T_1$
\ENSURE Optimized 3D sketch $\mathcal{S} = \{S_i\}_{i=1}^N$

\STATE Initialize 3D Bézier curves: $S_i = \{p_{i,j} \in \mathbb{R}^3\}_{j=0}^3$ randomly
\STATE Define canonical viewpoints: $\mathcal{V} = \{\text{front}, \text{back}, \text{left}, $
$\text{right}\}$

\FOR{$t = 1$ to $T_1$}
    \STATE $\mathcal{L}_{\text{total}} \leftarrow 0$
    
    \FOR{each view $v \in \mathcal{V}$}
        \STATE $I_v \leftarrow \text{DifferentiableRender}(\mathcal{S}, v)$ \COMMENT{Project 3D curves to 2D}
        \STATE $y_v \leftarrow \text{"A } v \text{ view of } y\text{"}$ \COMMENT{View-dependent prompting}
        \STATE $\mathcal{L}_v \leftarrow \text{SDS}(I_v, y_v)$ \COMMENT{Score Distillation Sampling}
        \STATE $\mathcal{L}_{\text{total}} \leftarrow \mathcal{L}_{\text{total}} + \mathcal{L}_v$
    \ENDFOR
    
    \STATE $\mathcal{L}_{\text{geom}} \leftarrow \text{GeometricConsistency}(\mathcal{S})$ \COMMENT{Structural regularization}
    \STATE $\mathcal{L}_{\text{stage1}} \leftarrow \mathcal{L}_{\text{total}} + \lambda_g \mathcal{L}_{\text{geom}}$
    
    \STATE $\{p_{i,j}\} \leftarrow \{p_{i,j}\} - \alpha \nabla_{p_{i,j}} \mathcal{L}_{\text{stage1}}$ \COMMENT{Update control points}
\ENDFOR

\RETURN $\mathcal{S}$
\end{algorithmic}
\end{algorithm}

\vspace{-10pt}
\subsection{Stage II: Motion Field Learning for 3D Animation}

\vspace{-10pt}
\begin{algorithm}
\caption{Stage II: Motion Field Learning for 3D Animation}
\begin{algorithmic}[1]
\REQUIRE Static 3D sketch $\mathcal{S}$, motion prompt $y_{\text{motion}}$, frames $K$, iterations $T_2$
\ENSURE Animated 3D sketch sequence $\{\mathcal{S}^{(k)}\}_{k=1}^K$

\STATE Initialize displacement field: $\Delta p_{i,j}^{(k)} = \mathbf{0}$ for all $i,j,k$

\FOR{$t = 1$ to $T_2$}
    \STATE \COMMENT{Project 3D motion to 2D views}
    \FOR{$k = 1$ to $K$}
        \STATE $p_{i,j}^{(k)} \leftarrow p_{i,j}^{(0)} + \Delta p_{i,j}^{(k)}$ \COMMENT{Apply displacement}
        
        \STATE \COMMENT{Front view projection (xy coordinates)}
        \STATE $v_{\text{front}}^{(k)} \leftarrow \text{flatten}(\{(x_{i,j}^{(k)}, y_{i,j}^{(k)})\}_{i,j})$
        
        \STATE \COMMENT{Side view projection (yz coordinates)}
        \STATE $v_{\text{side}}^{(k)} \leftarrow \text{flatten}(\{(y_{i,j}^{(k)}, z_{i,j}^{(k)})\}_{i,j})$
    \ENDFOR
    
    \STATE \COMMENT{Video SDS optimization}
    \STATE $\mathcal{L}_{\text{front}} \leftarrow \text{VideoSDS}(\{v_{\text{front}}^{(k)}\}_{k=1}^K, y_{\text{motion}})$
    \STATE $\mathcal{L}_{\text{side}} \leftarrow \text{VideoSDS}(\{v_{\text{side}}^{(k)}\}_{k=1}^K, y_{\text{motion}})$
    
    \STATE \COMMENT{Predict displacement vectors}
    \STATE $\{\Delta v_{\text{front}}^{(k)}\} \leftarrow \text{PredictDisplacement}(\mathcal{L}_{\text{front}})$
    \STATE $\{\Delta v_{\text{side}}^{(k)}\} \leftarrow \text{PredictDisplacement}(\mathcal{L}_{\text{side}})$
    
    \STATE \COMMENT{Reconstruct 3D displacement}
    \FOR{each curve $i$, control point $j$, frame $k$}
        \STATE $\Delta x_{i,j}^{(k)} \leftarrow \Delta v_{\text{front}}^{(k)}[2(i \cdot 4 + j)]$
        \STATE $\Delta y_{i,j}^{(k)} \leftarrow \frac{\Delta v_{\text{front}}^{(k)}[2(i \cdot 4 + j) + 1] + \Delta v_{\text{side}}^{(k)}[2(i \cdot 4 + j)]}{2}$
        \STATE $\Delta z_{i,j}^{(k)} \leftarrow \Delta v_{\text{side}}^{(k)}[2(i \cdot 4 + j) + 1]$
        \STATE $\Delta p_{i,j}^{(k)} \leftarrow [\Delta x_{i,j}^{(k)}, \Delta y_{i,j}^{(k)}, \Delta z_{i,j}^{(k)}]^T$
    \ENDFOR
    
    \STATE \COMMENT{Temporal smoothness regularization}
    \STATE $\mathcal{L}_{\text{smooth}} \leftarrow \sum_{i,j,k} \|\Delta p_{i,j}^{(k+1)} - \Delta p_{i,j}^{(k)}\|_2^2$
    \STATE $\mathcal{L}_{\text{motion}} \leftarrow \mathcal{L}_{\text{front}} + \mathcal{L}_{\text{side}} + \lambda_s \mathcal{L}_{\text{smooth}}$
    
    \STATE Update displacement field: $\{\Delta p_{i,j}^{(k)}\} \leftarrow \{\Delta p_{i,j}^{(k)}\} - \beta \nabla \mathcal{L}_{\text{motion}}$
\ENDFOR

\FOR{$k = 1$ to $K$}
    \STATE $\mathcal{S}^{(k)} \leftarrow \{p_{i,j}^{(0)} + \Delta p_{i,j}^{(k)}\}_{i,j}$ \COMMENT{Generate final frames}
\ENDFOR

\RETURN $\{\mathcal{S}^{(k)}\}_{k=1}^K$
\end{algorithmic}
\end{algorithm}

\section{Training Dynamics and Parameter Sensitivity Analysis}

This section examines the impact of key training parameters on generation quality, with Figure~\ref{fig:3D-animation1} and Figure~\ref{fig:3D-animation2} to~\ref{fig:3D-animation3} providing a comprehensive visualization of our generated 3D dynamic sketches. Figure~\ref{fig:3D-resulte1} to~\ref{fig:3D-result2} show more high quality 3D models generated by our method. Through systematic experimentation, we find that model convergence and generation performance are particularly sensitive to two parameters: classifier-free guidance (CFG) strength and the number of strokes.

The choice of CFG value directly influences convergence speed (Figure~\ref{fig:sup1}). Higher CFG values lead to faster convergence, indicating that stronger conditional guidance accelerates the training process. However, excessive CFG values can cause instability in generation quality, necessitating a balance between convergence speed and generation stability.

Regarding stroke count, our experiments reveal its significant impact on model convergence. As demonstrated in Figure~\ref{fig:sup1}, insufficient stroke counts impede the model from reaching optimal convergence. Conversely, with an appropriate stroke count (e.g., 32 strokes), the model achieves faster convergence. This finding not only provides practical guidance for stroke count selection but also motivates our investigation into the reliability of evaluation metrics.

\section{Limitations of Evaluation Metrics}

Current CLIP-based evaluation paradigms exhibit fundamental limitations in assessing 3D sketch generation, affecting both model evaluation and broader research development. Our systematic investigation reveals four critical limitations in CLIP scoring mechanisms (Figure~\ref{fig:sup2}).

\subsection{Inconsistency in Structure Expression}

We observe significant discrepancies between CLIP scores and structural expressiveness. With 16 strokes, despite unclear structural representation, the CLIP score reaches 0.3009. Increasing to 32 strokes improves structural richness, raising the score to 0.3288. However, further increasing to 64 strokes, despite enabling more refined details, paradoxically reduces the score to 0.3123. This indicates no simple positive correlation between CLIP scores and actual structural expressiveness.

\subsection{Quality Assessment Paradox}
\begin{figure}
    \centering
    \includegraphics[width=\linewidth]{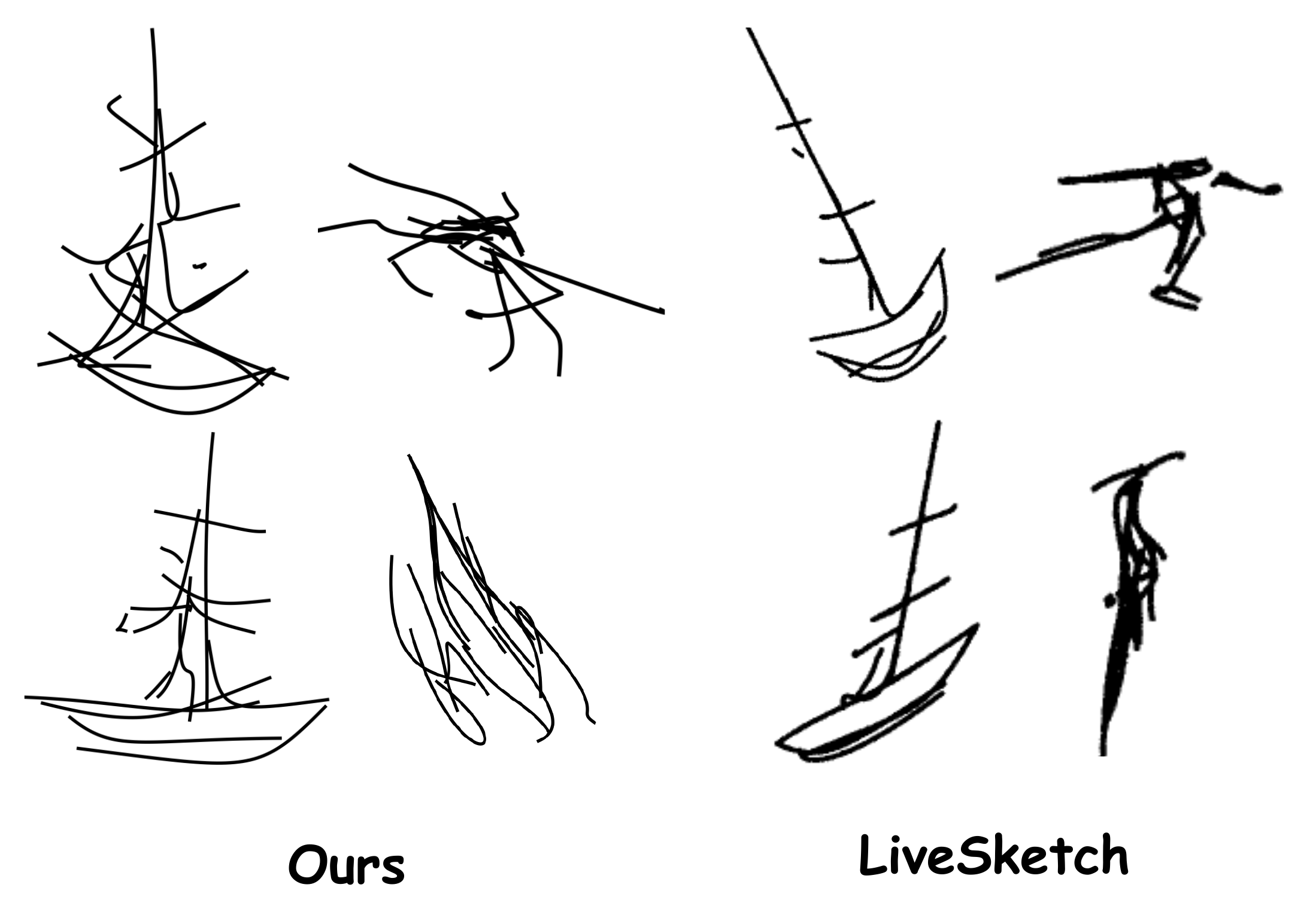}
    \caption{The issue of straightening lines in LiveSketch.}
    \label{fig:rigid-image}
\end{figure}
More critically, we observe significant bias in CLIP's quality assessment. In a comparative experiment with three fish sketches, a flawed sketch showing only half a fish body (Fish 1) achieves a high CLIP score of 0.3000. Conversely, a refined sketch (Fish 2) generated through carefully crafted prompts receives only 0.2847. Fish 3 is the result when using a normal SD-v2.1. This clearly demonstrates CLIP's tendency to overemphasize local features while neglecting overall artistic quality and structural coherence.

\begin{figure*}
    \centering
    \includegraphics[width=\linewidth]{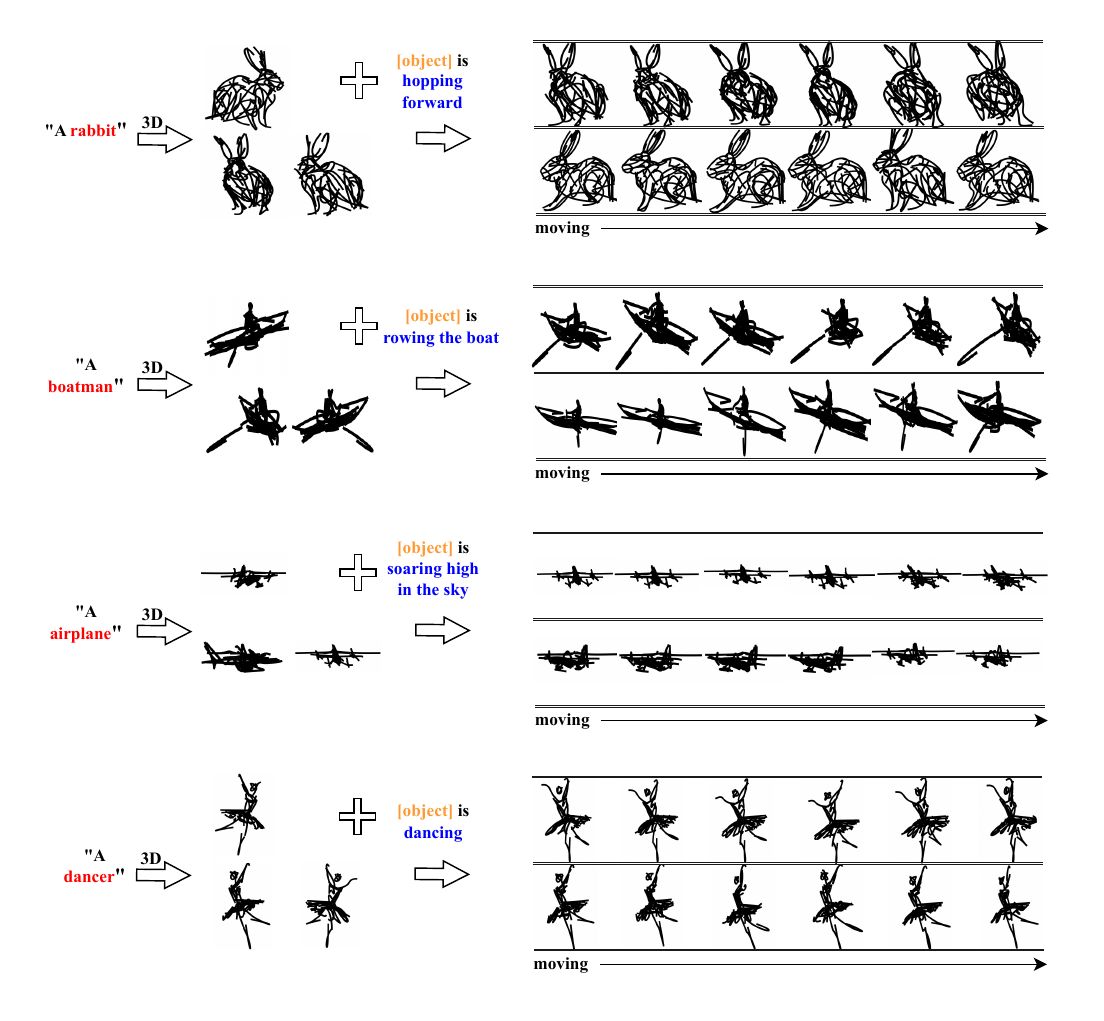}
    \caption{Rich 3D dynamic animation generated by our model.}
    \label{fig:3D-animation1}
\end{figure*}

\begin{figure*}
    \centering
    \includegraphics[width=\linewidth]{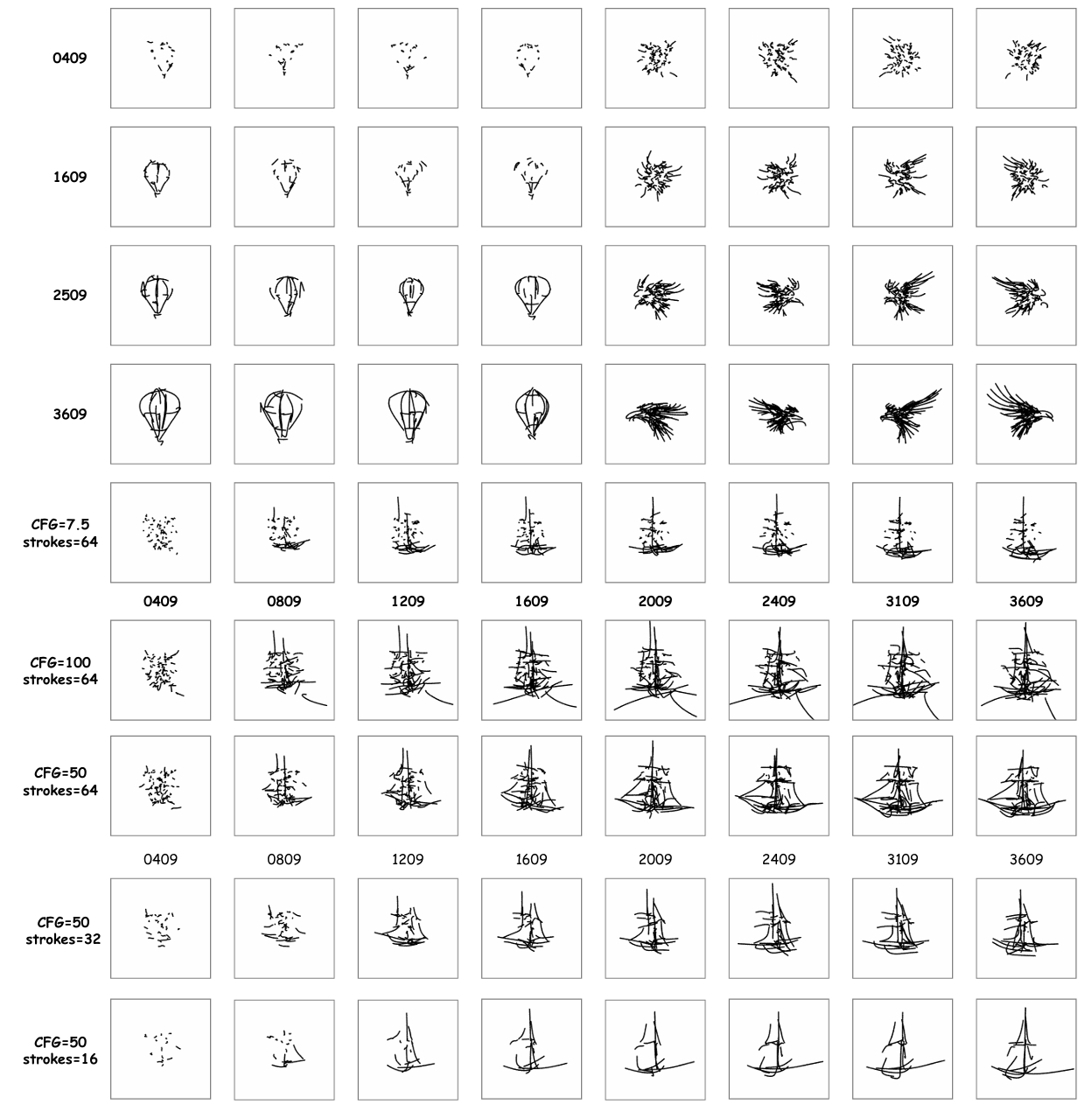}
    \caption{Parameter Sensitivity Analysis.}
    \label{fig:sup1}
\end{figure*}

\begin{figure*}
    \centering
    \includegraphics[width=\linewidth]{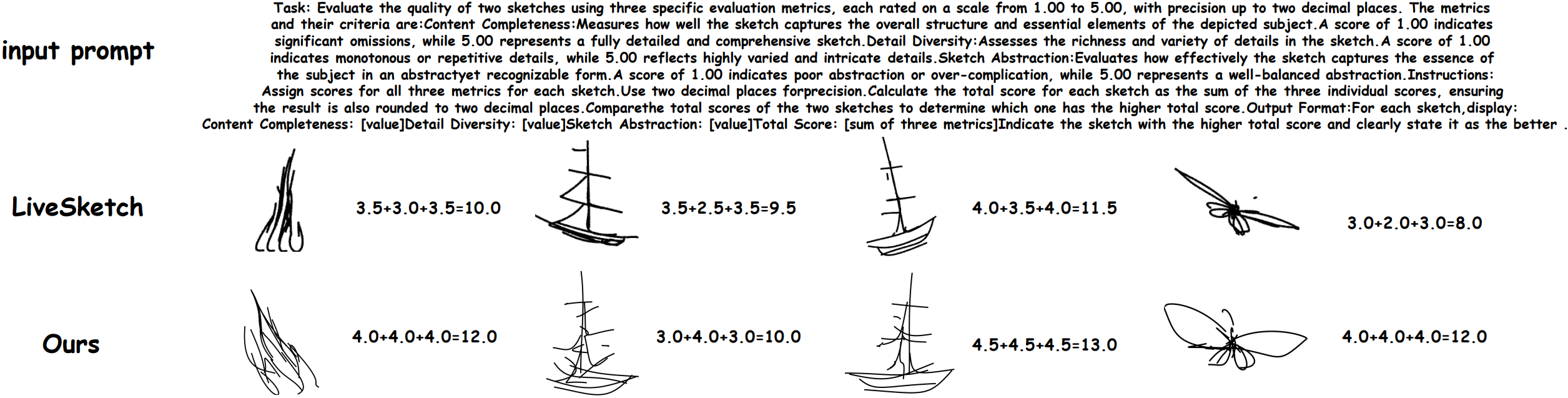}
    \caption{The evaluation value derived from Qwen-VLM.}
    \label{fig:qwen}
\end{figure*}

\begin{figure*}
    \centering
    \includegraphics[width=\linewidth]{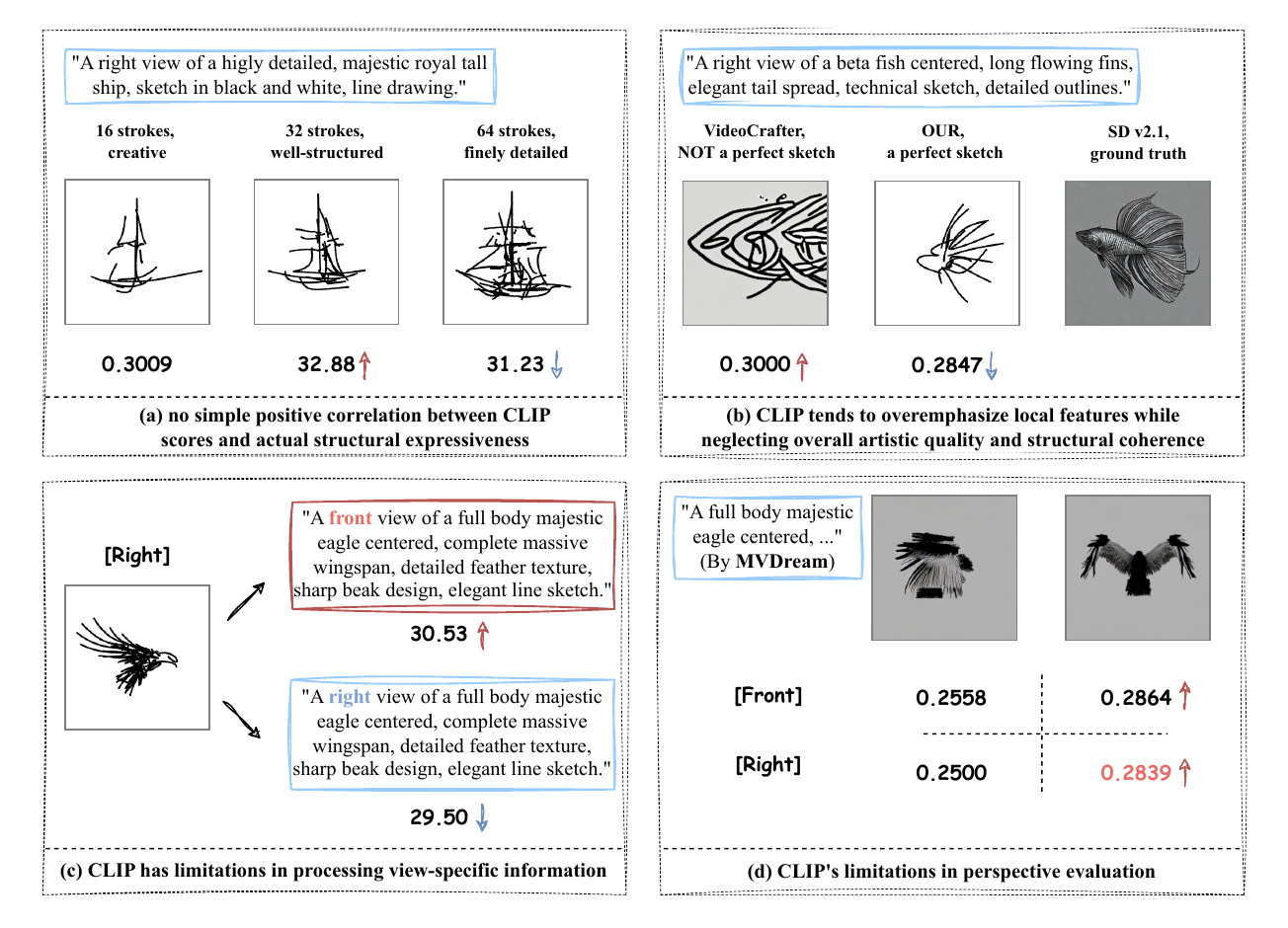}
    \caption{Four critical limitations in CLIP scoring mechanisms.}
    \label{fig:sup2}
\end{figure*}


\subsection{View-Specificity Limitations}

The third key finding concerns CLIP's view perception capabilities. Testing with a side-view eagle sketch, the prompt ``a front view" yields a score of 0.3053, while the accurate ``a left view" description scores lower at 0.2950. This counter-intuitive result directly evidences CLIP's limitations in processing view-specific information.

\subsection{Multi-View Consistency Deficiencies}

Further validation using MVDream's multi-view consistent images reinforces CLIP's view assessment limitations. While front views score appropriately higher with ``a front view of..." prompts, the model fails for side views: when prompted with ``a right view of...", front views still score higher (0.2839) than accurate side views (0.2500). This reveals systematic deficiencies in handling multi-view scenarios.

These experimental findings collectively point to a crucial conclusion: existing CLIP-based evaluation methods have fundamental limitations in assessing 3D sketch generation tasks, necessitating the development of more specialized and effective evaluation frameworks. This understanding drives our exploration of improved evaluation approaches, which will be discussed in detail in subsequent sections.

\section{Analysis of Cascaded Space Complementarity}

Our experimental observations reveal a significant phenomenon that validates the necessity of our cascaded complementary space design. When feeding our generated 3D sketch projections into the Live Sketch model for dynamic generation, we observe notable quality degradation. As shown in Figure~\ref{fig:rigid-image}, Live Sketch forcibly converts our rhythmic and creative strokes into rigid straight lines. This ``normalization" process essentially eliminates the artistic expressiveness inherent in the original sketches, directly impacting the naturalness of subsequent dynamic generation.


This observation highlights a crucial limitation in existing single-space approaches when handling artistically expressive sketches. Live Sketch's behavior largely stems from its training data bias, primarily using CLIPasso sketches that closely adhere to real image edges, lacking the artistic expressiveness of traditional hand-drawn sketches. This data preference leads to the rejection of non-regular strokes, limiting artistic quality and expressiveness in the generated results.

In contrast, our cascaded complementary space design establishes a solid foundation for dynamic generation during the static generation phase. By maintaining appropriate artistic freedom in the structure space, it enables more natural dynamic effects in the motion space. Figure~\ref{fig:rigid-image} demonstrates this advantage: our method preserves basic object structure while conveying richer visual information through expressive strokes. This directly evidences the importance of static generation quality for subsequent dynamic generation, validating our cascaded space design.

\section{Improved Evaluation Framework}

Addressing the identified limitations of CLIP scoring mechanisms, we propose a comprehensive evaluation framework leveraging Qwen2-VL-7B, an open-source vision-language model. Our framework implements a systematic scoring system with precisely defined metrics, each rated on a standardized scale from 1.00 to 5.00 with two decimal precision, ensuring quantitative rigor in assessment.


The evaluation framework centers on three fundamental dimensions: Content Completeness, Detail Diversity, and Sketch Abstraction. Content Completeness evaluates how comprehensively the sketch captures the overall structure and essential elements of the depicted subject, with scores ranging from significant omissions (1.00) to fully detailed representation (5.00). Detail Diversity measures the richness and variety of artistic elements, distinguishing between monotonous execution (1.00) and intricate, varied detailing (5.00). Sketch Abstraction assesses the delicate balance between abstraction and recognition, where lower scores indicate either overly simplified or unnecessarily complex representations, and higher scores reflect an optimal balance of abstract expression while maintaining recognizability.

\begin{figure*}
    \centering
    \includegraphics[width=\linewidth]{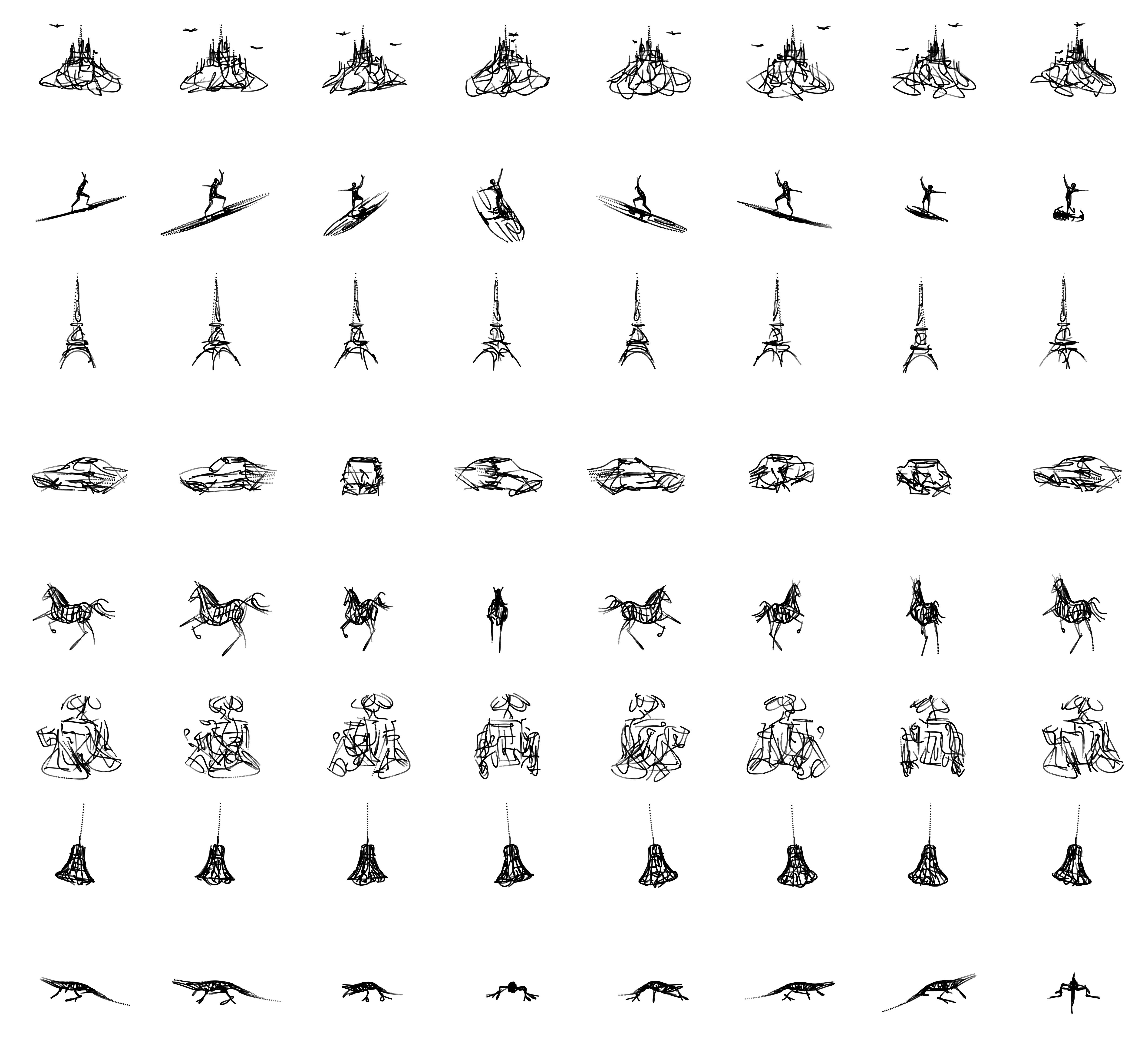}
    \caption{Rich 3D results generated by our model}
    \label{fig:3D-result2}
\end{figure*}
\begin{figure*}
    \centering
    \includegraphics[width=\linewidth]{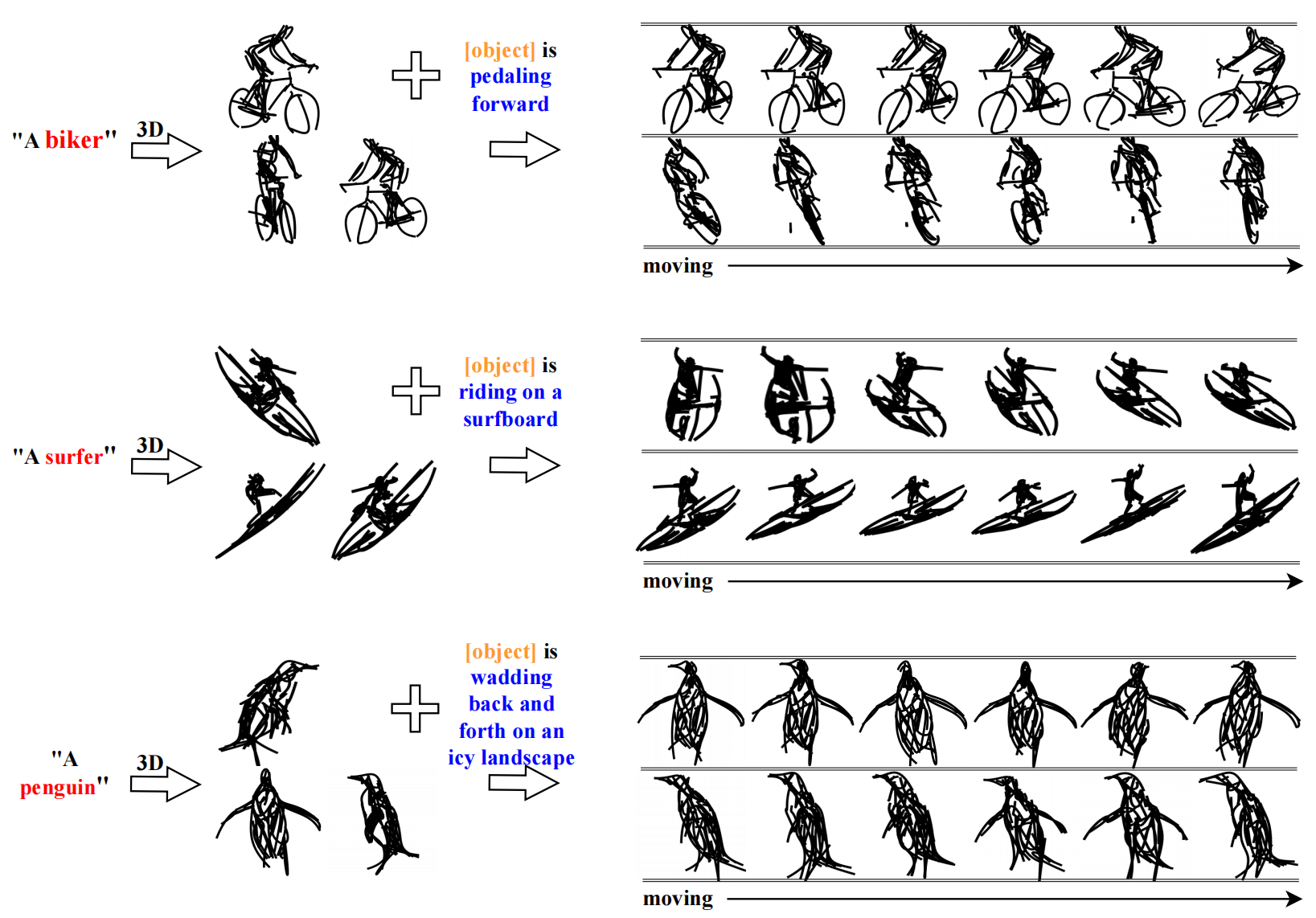}
    \caption{Rich 3D dynamic animation generated by our model.}
    \label{fig:3D-animation2}
\end{figure*}

The Qwen-VL model plays a central role in this evaluation process, employing carefully crafted prompt templates that guide the model to assess each dimension from an artistic evaluation perspective. As demonstrated in Figure~\ref{fig:qwen}, the evaluation yields not only quantitative scores but also provides detailed qualitative analysis, offering insights into the strengths and weaknesses of each generated sketch.


To ensure reliability and reproducibility in evaluation, our framework implements a comprehensive scoring system where the final assessment is derived from the sum of scores across all three dimensions. This approach maintains the distinct insights from each dimension while providing a single, comparable metric for overall sketch quality. The totality of scores, as shown in Figure~\ref{fig:qwen}, enables direct comparison between different sketches while preserving the nuanced evaluation of individual aspects.


Experimental results demonstrate that this multi-dimensional quantitative assessment approach not only overcomes the limitations of CLIP-based evaluation but also provides a more comprehensive and reliable standard for sketch quality assessment. The framework shows particular strength in evaluating abstract expression and detail handling, aspects that require deep artistic understanding and have traditionally been challenging to assess programmatically.

\begin{figure*}
    \centering
    \includegraphics[width=\linewidth]{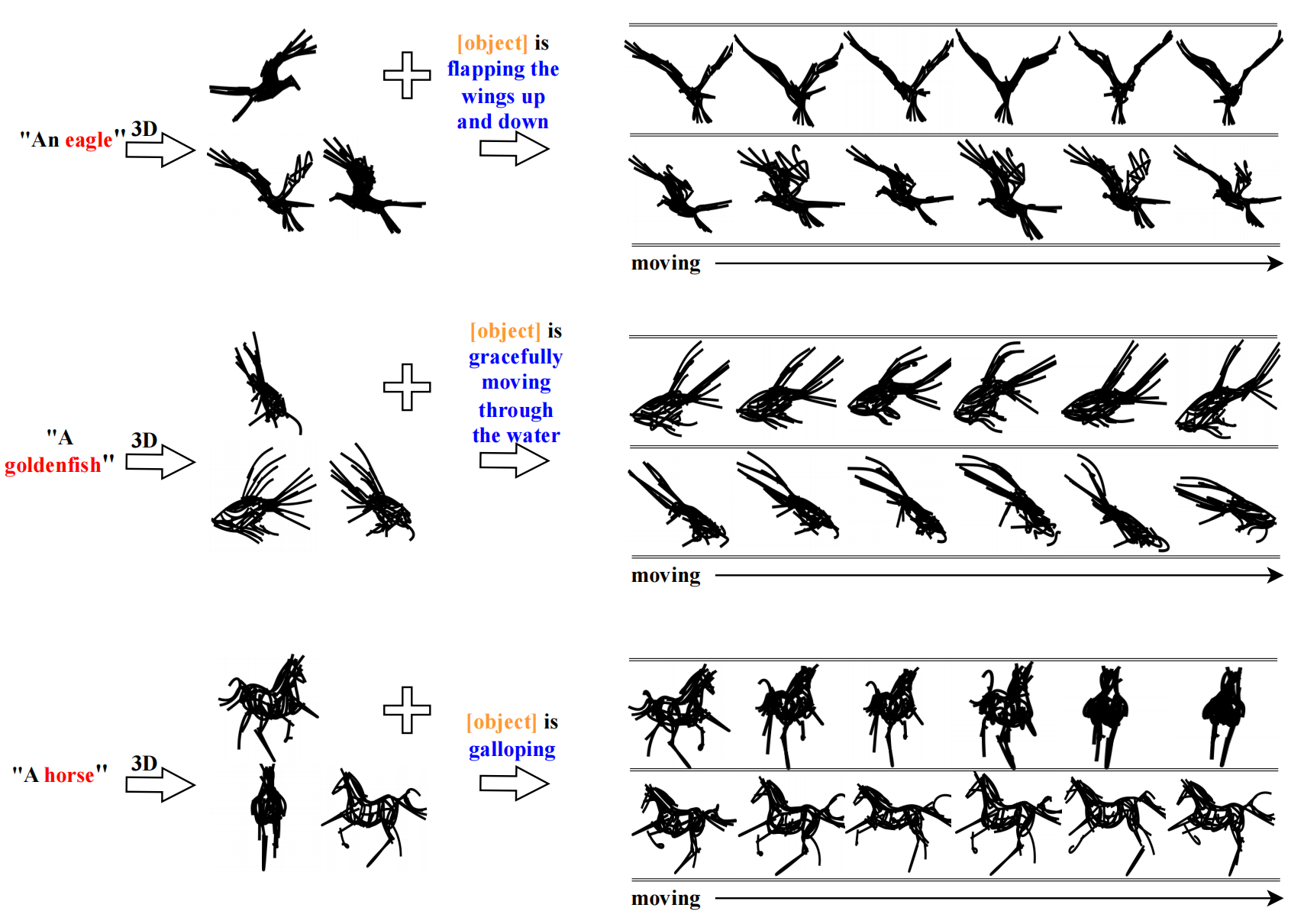}
    \caption{Rich 3D dynamic animation generated by our model.}
    \label{fig:3D-animation3}
\end{figure*}
\begin{figure*}
    \centering
    \includegraphics[width=\linewidth]{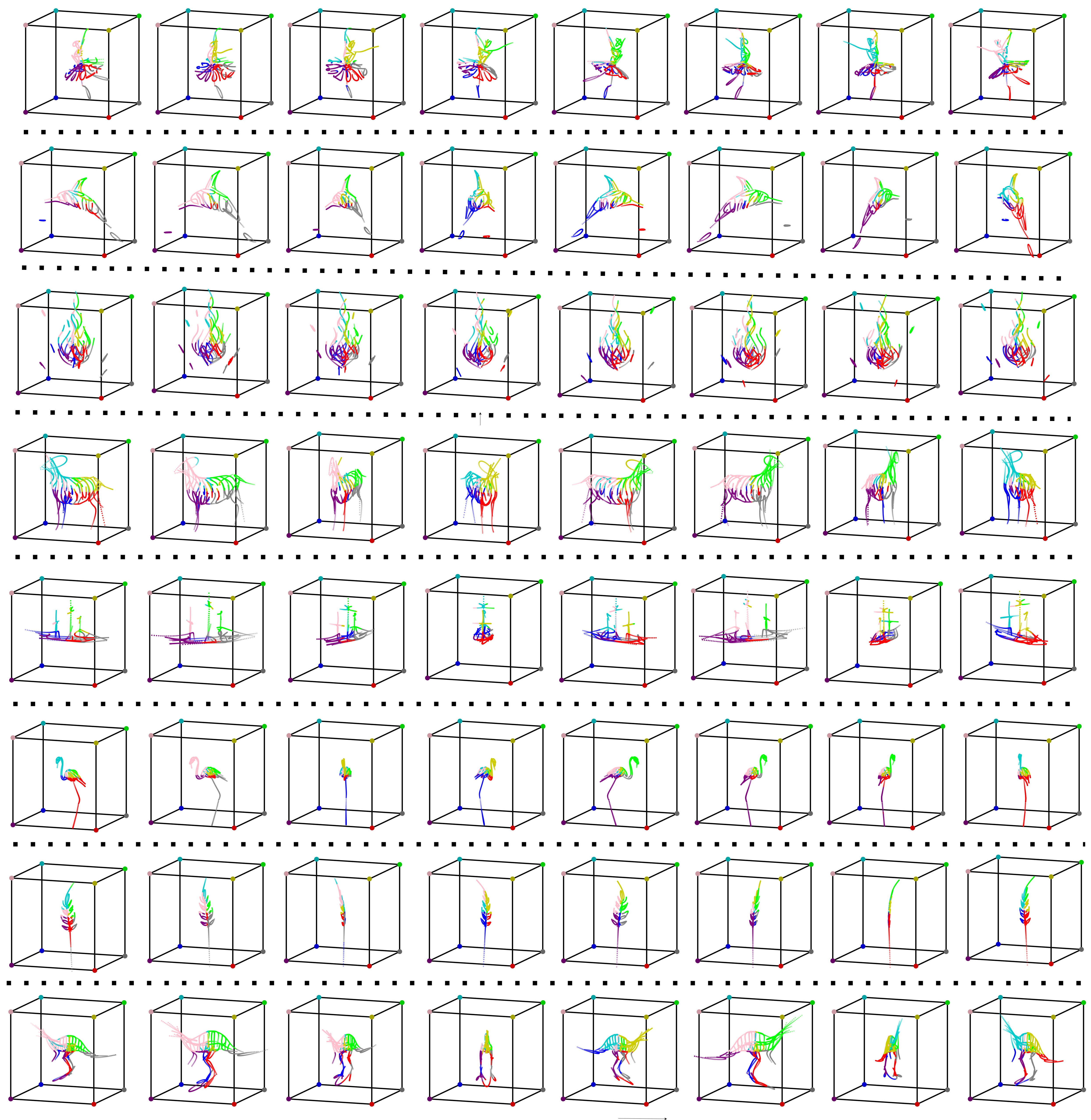}
    \caption{Rich 3D results generated by our model}.
    \label{fig:3D-resulte1}
\end{figure*}

\end{document}